%% file: main_TR.tex
\documentclass[pdflatex,sn-mathphys-num]{sn-jnl}

\usepackage{graphicx}%
\usepackage{multirow}%
\usepackage{amsmath,amssymb,amsfonts}%
\usepackage{amsthm}%
\usepackage[title]{appendix}%
\usepackage{xcolor}%
\usepackage{textcomp}%
\usepackage{manyfoot}%
\usepackage{booktabs}%
\usepackage{algorithm}%
\usepackage{algorithmicx}%
\usepackage{algpseudocode}%
\usepackage{listings}%
\usepackage{placeins}%

\usepackage{chemformula}

\theoremstyle{thmstyleone}%
\theoremstyle{thmstyletwo}%

\theoremstyle{thmstylethree}%
\newcommand{\ppcrn}{\textit{Popcornn}}
\newcommand{\pdq}{\textit{PadaQuad}}
\makeatletter\def\toclevel@subparagraph{1}\makeatother

\newcommand{\vremm}{\mathcal{L}_\text{VRE}}
\newcommand{\vre}{$\vremm$}
\newcommand{\pvremm}{\mathcal{L}_\text{pVRE}}
\newcommand{\pvre}{$\pvremm$}
\newcommand{\geomm}{\mathcal{L}_\text{geo}}
\newcommand{\geo}{$\geomm$}
\newcommand{\tslossmm}{\mathcal{L}_\text{TS}}
\newcommand{\tsloss}{$\tslossmm$}
\newcommand{\tsrlossmm}{\mathcal{L}_\text{TSR}}
\newcommand{\tsrloss}{$\tsrlossmm$}

\newcommand{\rxnmm}{\mathbf{x}_\theta(t)}
\newcommand{\rxn}{$\rxnmm$}
\newcommand{\velmm}{\mathbf{v}_\theta(t)}
\newcommand{\vel}{$\velmm$}
\newcommand{\xrmm}{\mathbf{x}_R}
\newcommand{\xr}{$\xrmm$}
\newcommand{\xpmm}{\mathbf{x}_P}
\newcommand{\xp}{$\xpmm$}
\newcommand{\fnnmm}{\mathbf{f}_\theta(t)}
\newcommand{\fnn}{$\fnnmm$}

\begin{document}
\title[Article Title]{Neural Network Path Optimization for Finding Transition States on a Machine Learned Potential}


\author*[1,2]{\fnm{Kareem} \sur{Hegazy}}\email{khegazy@berkeley.edu}\equalcont{These authors contributed equally to this work.}
\author[3,6]{\fnm{Eric C-Y} \sur{Yuan}}\equalcont{These authors contributed equally to this work.}
\author[7]{\fnm{Anup} \sur{Kumar}}
\author[8]{\fnm{Dmitriy} \sur{Morozov}}
\author[7]{\fnm{Daniel} \sur{Barter}}
\author[1,2,8]{\fnm{Michael W} \sur{Mahoney}}
\author*[3,4,5,6]{\fnm{Teresa} \sur{Head-Gordon}}\email{thg@berkeley.edu}
\author*[7]{\fnm{Samuel M} \sur{Blau}}\email{smblau@lbl.gov}

\affil[1]{\orgdiv{Department of Statistics}, \orgname{University of California}, \orgaddress{\city{Berkeley}, \postcode{94720}, \state{CA}, \country{USA}}}

\affil[2]{\orgname{International Computer Science Institute}, \orgaddress{\city{Berkeley}, \postcode{94704}, \state{CA}, \country{USA}}}

\affil[3]{\orgdiv{Kenneth S. Pitzer Theory Center and Department of Chemistry}, \orgname{University of California}, \orgaddress{\city{Berkeley}, \postcode{94720}, \state{CA}, \country{USA}}}

\affil[4]{\orgdiv{Department of Bioengineering}, \orgname{University of California}, \orgaddress{\city{Berkeley}, \postcode{94720}, \state{CA}, \country{USA}}}

\affil[5]{\orgdiv{Department of Chemical and Biomolecular Engineering}, \orgname{University of California}, \orgaddress{\city{Berkeley}, \postcode{94720}, \state{CA}, \country{USA}}}

\affil[6]{\orgdiv{Chemical Sciences Division}, \orgname{Lawrence Berkeley National Laboratory}, \orgaddress{\city{Berkeley}, \postcode{94720}, \state{CA}, \country{USA}}}

\affil[7]{\orgdiv{Energy Technologies Area}, \orgname{Lawrence Berkeley National Laboratory}, \orgaddress{\city{Berkeley}, \postcode{94720}, \state{CA}, \country{USA}}}

\affil[8]{\orgdiv{Computer Science Area}, \orgname{Lawrence Berkeley National Laboratory}, \orgaddress{\city{Berkeley}, \postcode{94720}, \state{CA}, \country{USA}}}


\abstract{\input{00_abstract}}


\maketitle
\section{Introduction}
\input{01_introduction}

\section{Results}
\input{02_results}

\section{Discussion}
\input{03_discussion}

\section{Methods\label{sec:methods}}
\input{04_methods}

\section*{Data availability}
\input{data_availability}

\section*{Code availability}
\input{code_availability}


\backmatter





\bmhead{Acknowledgements}
\input{acknowledgements}
\input{funding}


\bmhead{Competing interests}
The authors declare no competing interests.

\bibliography{bibliography}



\newpage
\begin{appendices}
\renewcommand{\appendixname}{Supplementary Note}
\renewcommand{\theHfigure}{supp.\thefigure}
\renewcommand{\theHtable}{supp.\thetable}
\renewcommand{\theHequation}{supp.\theequation}

\section{Neural Path Representation}
\input{S01_path_representation}

\section{Popcornn Optimization Losses}
\input{S02_loss}

\section{Popcornn Convergence}
\input{S03_convergence}

\section{\label{sec:curriculum}Curriculum Optimization and Popcornn GSM}
\input{S04_GSM}

\section{\label{app:pdq_parallelism}Parallelized Adaptive Numerical Integration and \pdq}

\input{S05_parallel_processing}

\section{Hyperparameter Screening}
\input{S06_neb}

\section{\ppcrn{} Algorithms and Pseudocode}
\input{S09_pseudocode}





\end{appendices}


\end{document}

%% file: 00_abstract.tex
Transition states govern chemical reaction rates and outcomes, yet locating them remains a central challenge in computational chemistry.
Double-ended search methods find transition states given reactant and product structures, but generally rely on discretized path representations developed for density functional theory, where potential energy surface evaluations are expensive.
Machine learned interatomic potentials (MLIPs) dramatically reduce this cost and leverage automatic differentiation to access higher-order derivatives efficiently.
We introduce \ppcrn, a framework to find transition states by optimizing continuous neural network paths atop an MLIP with a parallelized adaptive numerical integrator.
This combination eliminates path discretization artifacts, reduces hyperparameter sensitivity, and enables non-local optimization to overcome local features using second-order potential-energy-surface information.
\ppcrn{} consistently recovers more correct transition states than standard methods such as nudged elastic band and growing string, across MLIPs on thousands of organic reactions, establishing it as a robust, automation-ready method for high-throughput transition state search.
\ppcrn's infrastructure is flexible in path objective, integration strategy, and optimization algorithm, making it a practical tool and platform for investigating open questions in MLIP-based path~optimization.

%% file: 01_introduction.tex
The quantum mechanical potential energy surface (PES) describes how an atomistic system's energy changes during conformational rearrangement and chemical transformations. The most chemically pertinent areas of the PES are stationary points: minima define stable molecules, and first-order saddle points define transition states (TSs) that describe transformations of reactant molecules into product species. Accurately and efficiently locating the TSs on the PES is critical for understanding complex reactive combustion \cite{Johansson2018, marti2023comprehensive}, electrolyte decomposition \cite{Xie2021,Spotte-Smith2022,Barter2023,Spotte-Smith2023,Spotte-Smith2023b}, heterogeneous \cite{Garza2018} and homogeneous catalysis \cite{Arnold2020}, polymer transformations \cite{Demarteau2022}, and more.

PES minima can be reliably found with a variety of different local optimization algorithms using only iterative energy and gradient (1st order and quasi-second order) methods, starting from nearly any reasonable initial guess structure~\cite{schlegel2003, Schlegel2011}. In contrast, single-ended TS optimization depends heavily on the quality of the initial guess (i.e., how close the structure is to the desired TS), the optimization algorithm employed, and often the need to calculate Hessian (2nd order) information. TS optimization for chemical transformations is typically carried out with density functional theory (DFT). However, when Hessian (2nd order) information is used, DFT eventually becomes intractable for large systems due to quartic scaling with the number of basis functions. As automatically-differentiable machine learned interatomic potentials (MLIPs) progress towards being reliable DFT surrogates, improved accessibility of second-order information has enabled more robust single-ended TS optimization \cite{Yuan2024,burger2025shoothiphessianinteratomic,wander2024accessingnumericalenergyhessians,Wu2026}. 

Even with readily-available Hessians, single-ended TS optimization remains fundamentally constrained by the need for a reasonably good initial guess TS structure \cite{Yuan2024,Wu2026}. 
Experts can build such structures by hand, structural templates can be employed for specific reaction classes~\cite{VandeVijver2020}, and generative models can now propose TS structures from reactants and products~\cite{Duan2023,Kim2024,Duan2025,Galustian2025} or from reactants alone~\cite{Han2026nexus}.
However, these generative models are limited in chemical scope by their TS training data; currently, standard datasets consist of main-group reactions~\cite{Schreiner2022,Han2026nexus}.
Therefore, automated initial guess construction across the chemical diversity of complex reaction networks~\cite{Blau2021,Wen2023} remains an open challenge.

An alternative and more automation-friendly approach is double-ended TS search~\cite{Schlegel2011}, which locates the TS given reactant and product structures by reformulating the task as a path optimization. 
It is the natural tool for reaction networks built by enumerating candidate reactant--product pairs~\cite{Zhao2021,Blau2021,Barter2023}. 
The path can be thought of as an optimizable function \rxn{} which maps a point on a reaction coordinate or ``reaction time'' $t$ to a 3D structure. The path is constrained to return the reactant structure when $t = 0$ and the product structure when $t = 1$: $\mathbf{x}_\theta(0) = \mathbf{x}_R$ and $\mathbf{x}_\theta(1) = \mathbf{x}_P$. Thus, the transformation of the reactant to the product must necessarily be accomplished along the path.
The optimization objective for most double-ended path optimization schemes is derived from or can be interpreted as a path integral:
\begin{equation}
    \mathcal{L} = \int_0^1 \ell(\mathbf{x}(t)) \, dt ,
    \label{eq:loss_path_integral}
\end{equation}
where $\ell(\mathbf{x}(t))$ quantifies the local ``path quality'' at reaction time $t$. In the simplest case, $\ell(\mathbf{x}(t))$ can be the potential energy $E(\mathbf{x}(t))$, and the minimization objective $\mathcal{L}_\text{E}$ is the total energy along the path. Decades of work, including the nudged elastic band (NEB) \cite{Jonsson1998,Henkelman2000} and growing string method (GSM) \cite{Zimmerman2013,Zimmerman2013b,Zimmerman2015,Jafari2017}, have been dedicated to obtaining the minimum energy path (MEP) by minimizing $\mathcal{L}_\text{E}$. The highest energy point along the MEP is guaranteed to be a TS, and the MEP is the gradient descent path from the TS to the reactant and product minima.

Notably, NEB and GSM were developed under the constraints of performing path optimization atop a DFT PES. Despite avoiding prohibitively expensive Hessians, calculating a structure's total energy and atomic forces with DFT still scales as $O(N_{e-}^3)$, becoming computationally challenging or intractable for larger systems.
Traditionally, to make the path optimization computationally practical, the path (and its integral Eq.~\ref{eq:loss_path_integral}) is discretized to a set of discrete structures (images) such that only tens of DFT calculations are needed at each optimization iteration.
However, path discretization has a variety of negative consequences: the potential to miss PES features between discrete images; complexities around how to place images along the path; maintaining path connectivity with fictitious spring forces; the need to treat a ``climbing image'' differently from the rest of the path; and the introduction of many hyperparameters (related to these details) that often need to be carefully selected by the user for each reaction.
All of these consequences are anathema to automation. Perhaps most problematic, path optimization proceeds locally such that each image's position is updated pursuant only to the energy and forces at that image and its two neighboring images, preventing the path from cohesively mutating into an entirely different track along the PES. 

Independent of the path representation, the MEP optimization objective is critically hindered by the MEP's sensitivity to local structures that do not affect the TS. Although $\mathcal{L}_\text{E}$ only requires PES first derivatives at each optimization iteration and the MEP has attractive mathematical properties~\cite{Fukui1981,Sheppard2008}, it is also sensitive to every part of the path. This sensitivity is a crucial drawback, making the optimization landscape substantially more complex and difficult to navigate. In practice, the primary goal of path optimization should be to locate the TS. With the TS in hand, we can then obtain the MEP and/or confirm the associated endpoints via intrinsic reaction coordinate (IRC) or quasi-IRC~\cite{Fukui1981,Goodman2003} calculations. However, directly targeting the TS -- rather than the MEP -- requires a force-based path objective~\cite{Birkholz2015}; but a force-based path objective necessitates the calculation of second derivatives along the path at each iteration. While intractable with DFT, this presents an opportunity for improved path optimization methodology atop differentiable MLIPs. 

In this work, we present a software platform for MLIP-based double-ended TS search that we call \ppcrn: ``Path Optimization with a Continuous Representation Neural Network.''
\ppcrn{} leverages a neural network to parameterize the continuous path from reactant to product, combined with a GPU-parallelized adaptive numerical path integrator to calculate the path objective -- a critical combination we find necessary to ensure comprehensive PES sampling and to avoid many issues of traditional path discretization. This combination further facilitates global and cohesive path optimization by differentiating the path objective through the MLIP with respect to the path's weights and biases.
By default, \ppcrn{} employs a force-based optimization objective that directly targets the TS rather than seeking the MEP.
We demonstrate that, with fixed default settings, \ppcrn{} consistently achieves improved robustness and recovery of intended TSs, when compared with NEB and GSM, across multiple MLIPs and thousands of in-distribution and out-of-distribution test reactions.
Additionally, \ppcrn{} is engineered to be flexible with respect to path objective, numerical integration strategy, optimization algorithm, and path initialization strategy.
Thus, it is both of immediate practical utility for double-ended TS search atop a pre-trained MLIP, and it is also envisioned as a powerful platform for subsequent investigation and further improvement of MLIP-based path optimization methodologies.

%% file: 02_results.tex
\subsection{\ppcrn{} framework}

\begin{figure}[htb!]
    \centering
    \includegraphics[width=\linewidth]{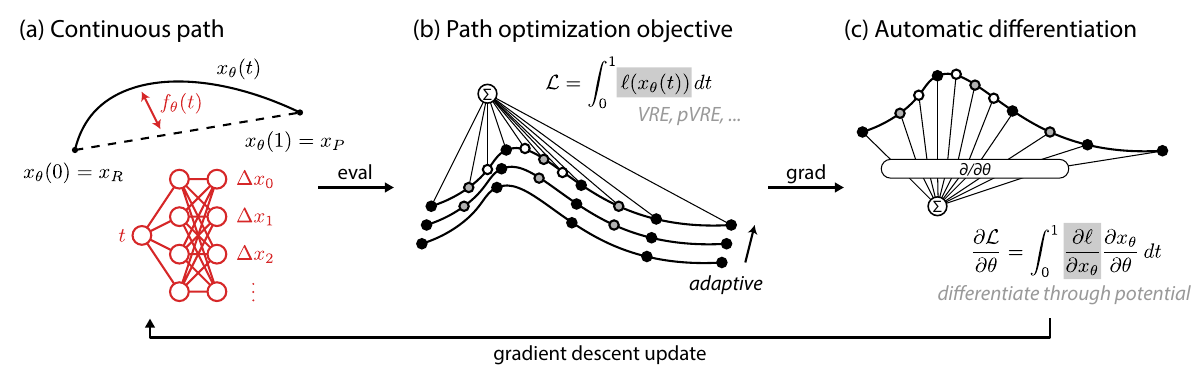}
    \caption{\textit{Schematic illustration of \ppcrn.} (a) The continuous reaction path is parameterized by a simple neural network. (b) The path objective of choice is calculated via a GPU-parallelized adaptive numerical path integrator. (c) The gradient of the objective with respect to the path neural network parameters is obtained via backpropagation through the differentiable potential, enabling path optimization via gradient descent.}
    \label{fig:model}
\end{figure}

\ppcrn{} uses a simple fully-connected neural network to parameterize the path, 
\begin{equation}
    \rxnmm = \underbrace{(1 - t) \, \xrmm + t \, \xpmm}_\text{linear prior} + \underbrace{t \, (1-t) \, \fnnmm}_\text{perturbation} ,
    \label{eq:path}
\end{equation}
where $t \in [0, 1]$ is the reaction time, \xr{} and \xp{} are the reactant and product endpoints, and \fnn{} is a fully-connected neural network with adjustable parameters $\theta$ (Fig.~\ref{fig:model}a).
The path ($\rxnmm \in \mathbb{R}^{3N}$) and neural network ($\fnnmm \in \mathbb{R}^{3N}$) model the positions of all $N$ atoms.
The path formulation in Eq.~\ref{eq:path} employs a perturbation from a linear interpolation between \xr{} and \xp{}, while enforcing hard endpoint constraints: $\mathbf{x}_\theta(0)=\xrmm$ and $\mathbf{x}_\theta(1)=\xpmm$.
We observe that the perturbative correction to a linear prior is important for path quality and optimization speed; see Supplementary Note~\ref{app:path} for details.
This path formulation was developed independently and concurrently in references \cite{Han2024,Ramakrishnan2025,Nam2025}; additionally, neural path models have also been trained across many reactions to predict paths directly, from the endpoints~\cite{Nam2025} or outward from a TS~\cite{Han2026nexus}, whereas \ppcrn{} optimizes each path on the target PES.
The continuous path allows one to probe the molecular structure at any point, guaranteeing no discontinuities and allowing all relevant PES features along the path to be sampled during optimization. 
Critically, we note that changing any internal neuron parameter in $\rxnmm$ modifies all atomic positions across the entire reaction path, enabling non-local path mutation rather than local bead perturbations. 

To realize our continuous formulation's (Eqs.~\ref{eq:loss_path_integral} and \ref{eq:path}) full capabilities, the loss must also be evaluated via adaptive numerical integration, rather than a coarse or other rigid summation scheme.
Thus, unlike conventional methods that approximate the loss (Eq.~\ref{eq:loss_path_integral}) using discrete points, we evaluate the path integral with the GPU-parallelized adaptive numerical path integrator \pdq~\cite{Hegazy2025} (Fig.~\ref{fig:model}b).
\pdq{} globally updates the entire path by integrating the gradient of the loss with respect to all network parameters along the entire path (Eq.~\ref{eq:grad_path_integral}), driving cohesive path optimization via gradient descent (Fig.~\ref{fig:model}c).
The adaptivity of \pdq{} is central to avoiding fixed discretization, and it focuses PES evaluations in regions of high variation and large contributions to the integral while limiting evaluations in slowly varying regions (Supplementary Note~\ref{app:pdq_parallelism}).
The adaptive numerical integration of Eq.~\ref{eq:loss_path_integral} precludes the need for fictitious spring force terms.
Instead, \pdq{} does this by adaptively adding, removing, and adjusting evaluation points based on the integral error, independent of the PES value.
This adaptivity places evaluations where needed, removing the need for spring forces to affix them and creating an efficient and error-bound evaluation of the path.
We provide more details on \pdq{} in Supplementary Note~\ref{app:pdq_parallelism}.

We explore two different path objectives: one which targets the minimum energy path, and the other which directly targets the TS via the path which goes least high on the PES.
Crehuet and Bofill~\cite{Crehuet2005} introduced the variational reaction energy (VRE) path objective
\begin{equation}
    \vremm = \int_{0}^{1} | \mathbf{F}(\mathbf{x}_\theta(t))| \, |\mathbf{v}_\theta(t)| \, dt ,\label{eq:vre}
\end{equation}
where $\mathbf{F}(\rxnmm) \in \mathbb{R}^{3N}$ is the atomic forces at positions \rxn, and $\mathbf{v}_\theta(t)= d\rxnmm/dt \in \mathbb{R}^{3N}$ is the tangent to the path.
Minimizing \vre{} results in a minimum energy path, also known as a gradient descent path \cite{Aguilar-Mogas2007,Quapp2008}. This MEP is the goal of NEB and GSM.
Birkholz and Schlegel~\cite{Birkholz2015} introduced the projected VRE (pVRE) loss as a variational bound of \vre
\begin{equation}
    \pvremm = \int_{0}^{1} | \mathbf{F}(\mathbf{x}_\theta(t)) \cdot \mathbf{v}_\theta(t) | \, dt \label{eq:pvre} = \sum_a 2E_{a,TS} - E_{a,R} - E_{a,P} ,
\end{equation}
which is sensitive to the angle between $\mathbf{F}(\rxnmm)$ and $\mathbf{v}_\theta(t)$, and which can be thought of as a sum over all activation barriers $a$.
Minimizing \pvre{} results in a path that has the lowest PES barrier, thereby directly targeting the TS while being insensitive to structures along the rest of the path.
Later, we show \pvre{} produces qualitatively simpler and more robust paths.

The path \rxn{} is updated by the gradient of the path integral loss given by
\begin{equation}
    \frac{\partial\mathcal{L}}{\partial \theta} = \int_0^1 \frac{\partial \ell(t)}{\partial \theta} \, dt = \int_0^1 \frac{\partial \ell(\rxnmm)}{\partial \mathbf{x}_\theta} \frac{\partial \rxnmm}{\partial \theta} \, dt .
    \label{eq:grad_path_integral}
\end{equation}
Numerically, this gradient can be calculated in two ways: taking the gradient of the completed integral; or integrating the gradient of the integrand.
In the former case, the evaluation points are determined by the loss integral since the gradient is taken after.
In the latter case, the gradient is more accurate because the evaluation points are chosen to minimize the error of the gradient by placing them where the gradient of the loss varies the most.
We use the latter formulation, though \ppcrn{} and \pdq{} support both options.
The gradient is one order higher than the highest order derivative in the quantity $\ell$.
That is, if the integrand $\ell$ contains only the potential energy (like in NEB), only the first derivative of the PES (atomic forces) would be needed.
On the other hand, if $\ell$ involves the forces (VRE and pVRE), then the second derivative (the Hessian) is necessary for the path differentiation. The steep computational cost of DFT-based second derivatives has prevented force-based path objectives from being realistically viable, until the recent advent of differentiable MLIPs that are sufficiently accurate to describe reactive PESs for chemical systems of interest \cite{Yuan2024}.

The \pvre{} loss admits a natural force-based convergence criterion based on the optimization gradient (Eq.~\ref{eq:grad_path_integral}).
Established energy-minimization methods like NEB and GSM leverage modified forces (perpendicular gradients on the images and spring forces) to determine the convergence of the optimization~\cite{Henkelman2000}.
The \pvre{} loss isolates and targets the TS  (Supplementary Note~\ref{app:pvre}), while removing intermediate extrema along \rxn. 
Specifically, the gradient of \pvre{} becomes proportional to the TS force as the path converges (Eq.~\ref{eq:pvre_grad}).
Therefore, the gradient of \pvre{} is a physically consistent counterpart to convergence criteria used by the aforementioned established methods.
Given this correspondence, we set the same convergence criteria for \ppcrn, NEB, and GSM for the following chemical reaction evaluations.


\begin{figure}[htb!]
    \centering
    \includegraphics[width=\linewidth]{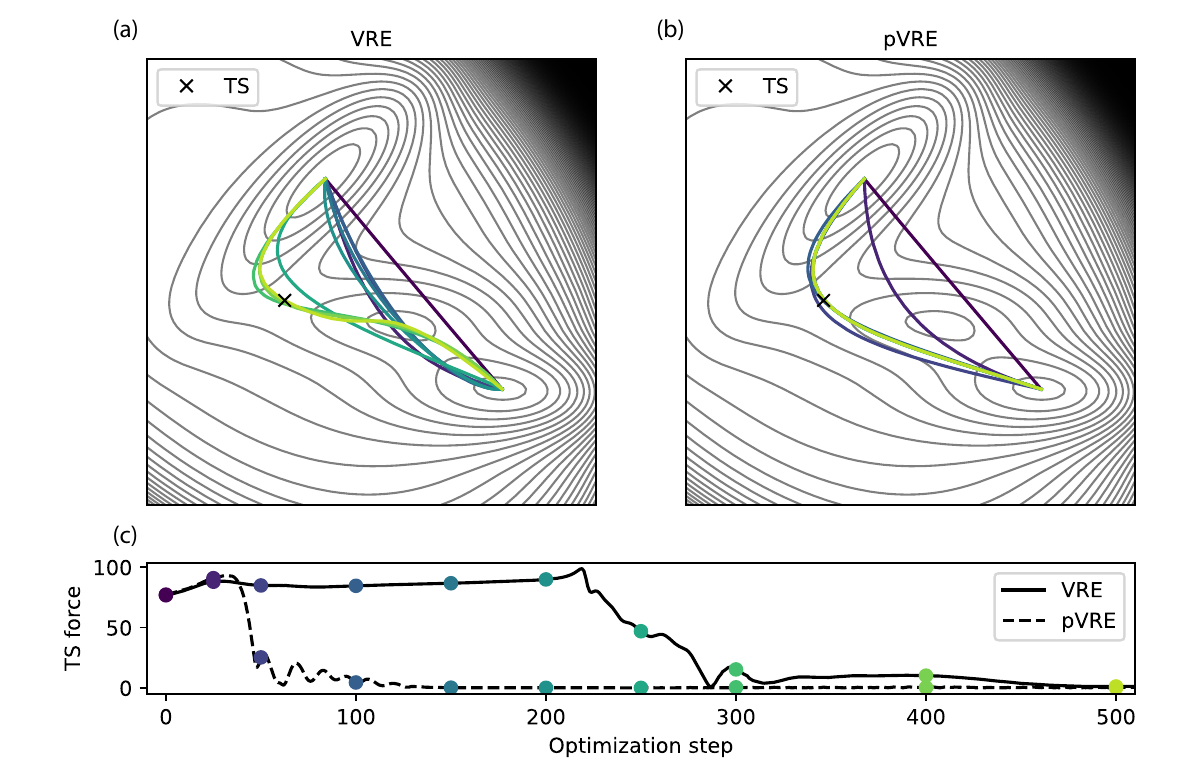}
    \caption{\textit{Different loss functions on the Muller-Brown potential.} (a) The VRE loss tries to find the minimum energy path, which can hinder efficient TS finding. (b) The pVRE loss focuses on minimizing the barrier height, directly targeting the TS. (c) The pVRE loss fully converges within 150 steps, whereas the VRE loss takes 500 steps to converge.}
    \label{fig:MB_vre_pvre}
\end{figure}


To understand the differences between paths optimized with the \pvre{} and \vre{} losses, we demonstrate \ppcrn{} with both loss functions on the model Müller-Brown potential, revealing the advantages of \pvre{}.
Figure~\ref{fig:MB_vre_pvre} shows the \ppcrn{} optimization of \rxn{} on the Müller-Brown potential for the \vre{} and \pvre{} losses.
We observe \rxn{} optimized with \pvre{} converges to the TS roughly three times faster than with \vre{}.
The \pvre{} optimization quickly locates the TS and converges to a stable path, while the \vre{} optimization struggles to navigate the PES within the first 200 steps due to \vre{}'s more complex locally-aware loss landscape. Given that our stated goal is to locate the TS as robustly as possible, the TS-focused \pvre{} is the superior path objective.
For the remainder of this work, we use \pvre{} as our optimization objective unless otherwise stated.
We provide more detailed discussion and intuition of the \vre{} and \pvre{} losses in Supplementary Note~\ref{app:loss}.

It is well established that the quality of the initial path provided for path optimization dramatically impacts the success rate of the path optimization task and the likelihood of recovering the intended TS~\cite{Rong2016}.
In the molecular context, the geodesic method of Zhu et al.~\cite{Zhu2019} and the IDPP method~\cite{Smidstrup2014} are commonly used. 
Both leverage discretized path optimizations with custom objectives atop bespoke analytic pairwise potentials to improve upon a naive linear path. This avoids steric clashes that could yield critically problematic DFT convergence or erroneous MLIP predictions.
\ppcrn{} extends its continuous path representation, adaptive numerical integration, and backpropagation-optimization machinery to initial path construction.
Similar to MLIPs, one can add new path initializations to the \ppcrn{} ecosystem~\cite{hegazy2026popcornn}.
We implemented a continuous path geodesic optimization atop the pairwise potential given by Zhu et al.~\cite{Zhu2019}, and we refer to this approach as \ppcrn-geodesic.
As we show below, this substantially improves path optimization outcomes.

\subsection{Chemical reaction evaluations}

\begin{figure}[H]
    \centering
    \includegraphics[width=\linewidth]{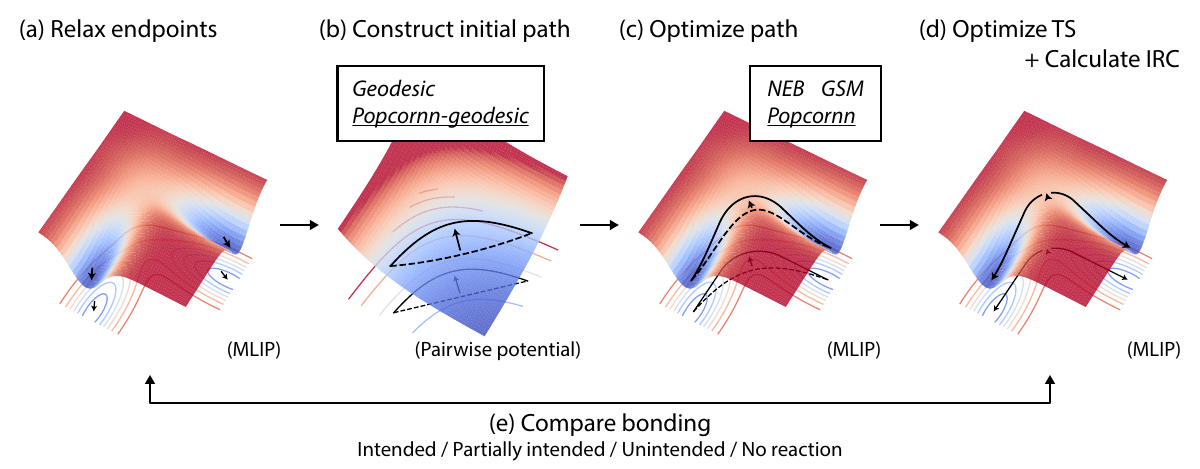}
    \caption{\textit{Systematic evaluation of double-ended TS finding methods.} (a) The endpoints are individually optimized to PES minima. (b) An initial path connecting the initial optimized endpoints is constructed via optimization atop a pairwise potential to approximate the actual reaction path and resolve steric clashes (Geodesic, \ppcrn-geodesic). (c) Path optimization is performed on the MLIP PES, starting from the initial path from (b) for NEB and \ppcrn, or directly from the optimized endpoints for GSM. (d) The highest point on the optimized path undergoes a single-ended TS optimization and is connected to the reaction endpoints using an IRC. (e) The IRC endpoints are compared with the initial endpoints to check how well the optimized TS corresponds to the intended reaction TS.}
    \label{fig:workflow}
\end{figure}

To evaluate \ppcrn's ability to robustly locate intended transition states in a systematic manner, we conduct large-scale experiments across different MLIPs  on both in-distribution and out-of-distribution reactions. By leveraging reaction test sets containing thousands of examples, we seek to ensure that we are drawing reliable, statistically significant conclusions, while demonstrating \ppcrn's robustness and high-throughput capabilities. 
We find the reaction path in three steps: we first optimize the endpoint structures on the MLIP (Fig. \ref{fig:workflow}a); secondly, we construct an initial path (between the optimized endpoints) atop a pairwise potential using discretized geodesic or \ppcrn-geodesic (Fig. \ref{fig:workflow}b); finally, we optimize the path atop the MLIP (Fig. \ref{fig:workflow}c).
To find the TS, we apply a TS optimization method to the structure along the optimized path with maximal energy and minimal forces.
To validate the proposed TSs, we apply (quasi)IRC calculations to recover endpoints (Fig. \ref{fig:workflow}d) that are then compared with initial endpoints (Fig. \ref{fig:workflow}e).
Experimental success, as measured by graph matching between initial endpoints and IRC endpoints, is categorized by a quality hierarchy: 
\begin{itemize}
    \item Intended: Both IRC endpoints match both initial endpoints.
    \item Partially intended: One IRC endpoint matches one initial endpoint.
    \item Unintended: Neither IRC endpoint matches either initial endpoint; however, the IRC endpoints also do not match each other, i.e., a reaction has still occurred. 
    \item No reaction: IRC endpoints match, i.e., the transition state found corresponds only to a conformational change, not a reaction.
\end{itemize}

We benchmark \ppcrn{} against the two most common path optimzation methods, NEB \cite{Jonsson1998,Henkelman2000} and GSM \cite{Zimmerman2013,Zimmerman2013b,Zimmerman2015,Jafari2017}. For NEB, we compare two different initial path construction strategies: discretized geodesic and \ppcrn-geodesic. For GSM, initial path construction is not required; the path is built concurrently with optimization. For \ppcrn, \ppcrn-geodesic constructs the initial path prior to \ppcrn{} optimization atop the MLIP with the pVRE optimization objective. We note that for each method we run all experiments with multiple different sets of hyperparameters and present results from only the single best-performing set. This procedure is most representative of performance in a simple high-throughput context, yielding a fair comparison between methods. Further discussion of NEB hyperparameters is provided in Section~\ref{sec:reference_methods}.
Supplementary Note~\ref{app:neb} presents ablation tests across bead density and path initialization methods for NEB.

We perform experiments on small organic reactions with MLIPs trained on Transition-1x \cite{Schreiner2022} (T1x), and we test on out-of-distribution (OOD) reactions present in the Grambow-Green dataset (GG) ~\cite{Grambow2020} but absent in T1x.
The reactions in T1x are derived from GG, but in some cases DFT NEB fails to find the path of the GG reaction, and that reaction is excluded from T1x.
These challenging 1,342 reactions where DFT NEB fails make up our OOD test dataset.
When comparing against NEB, we use standard settings of 10 and 20 beads, along with an exhaustive 100 bead optimization.
The 100 bead optimization roughly matches the number of \ppcrn{} evaluations, thus extending NEB towards the numerical integration regime for a more direct comparison with \ppcrn{}.
We additionally tested NEB with 1000 beads, but we found significantly worse behavior.

Our first experiment compares \ppcrn{} and NEB performance on small organic reactions using the NewtonNet MLIP \cite{Haghighatlari2022} trained on T1x \cite{Schreiner2022} from Ref.~\cite{Yuan2024}. 
We validate each test reaction as having a true TS on the NewtonNet PES by running single-ended TS optimizations from the corresponding TS structure in the Grambow-Green dataset, followed by IRC calculations that must recover endpoints with the intended bonding.
This ensures that any result less than ``Intended'' is solely due to a failure of the double-ended TS search procedure, and not incompleteness of the MLIP energy surface.

\begin{figure}[htb!]
    \centering
    \includegraphics[width=\linewidth]{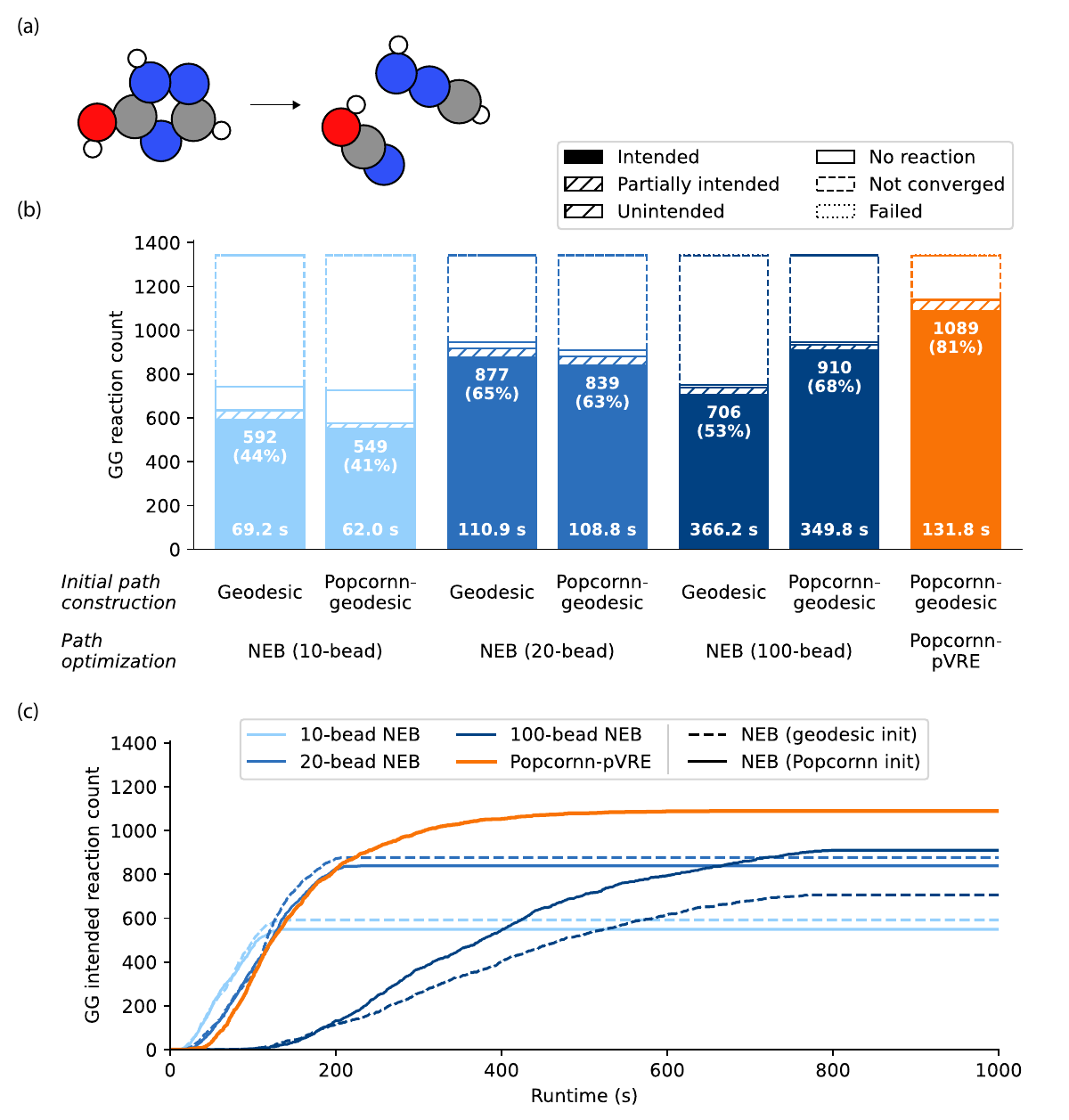}
    \caption{\textit{\ppcrn{} vs. NEB on small organic reactions using NewtonNet trained on Transition-1x.} (a) An example of the out-of-distribution reactions, those in the Grambow-Green dataset, on which we are testing. (b) \ppcrn{} finds more ``Intended'' TSs across various NEB setups with different initializations (Geodesic and \ppcrn-Geodesic) and number of beads (10, 20, 100). The number at the bottom of each bar is the average time, in seconds, for the TS optimization to converge. (c) \ppcrn's convergence rate to correct TSs is competitive with low-bead NEB for short runtimes, and ultimately outperforms all NEB variants.}
    \label{fig:bars_gg}
\end{figure}

\ppcrn-geodesic initial path construction followed by \ppcrn-pVRE atop the MLIP yields superior TS recovery over NEB with either initial path construction procedure.
\ppcrn{} achieves 81\% ``Intended'' TSs (Fig.~\ref{fig:bars_gg}b, right bar) compared with 44\%, 65\%, and 53\% for discrete geodesic $\rightarrow$ NEB with 10, 20, and 100 beads, respectively (Fig.~\ref{fig:bars_gg}b, left bars).
Interestingly, the \ppcrn-geodesic initial path improves over the discrete geodesic by 15\% for 100 NEB beads, and slightly worsens for 10 and 20 beads.
This indicates that a continuous or many-bead path benefits significantly from continuous optimization of the initial path. In contrast, starkly discretized path optimizations are not significantly impacted by a continuous initial path.
In addition to being more robust at convergence, \ppcrn{} is comparable in speed to both 10 and 20 bead NEB optimizations within the same accuracy range: roughly aligning with the lighter dashed blue lines before 200~s (Fig.~\ref{fig:bars_gg}c).
\ppcrn{} is significantly faster than the 100 bead path optimization.
Further comparison of alternative initial path construction strategies for NEB is provided in Supplementary Note~\ref{app:neb}.

\begin{figure}[tb!]
    \centering
    \includegraphics[width=\linewidth]{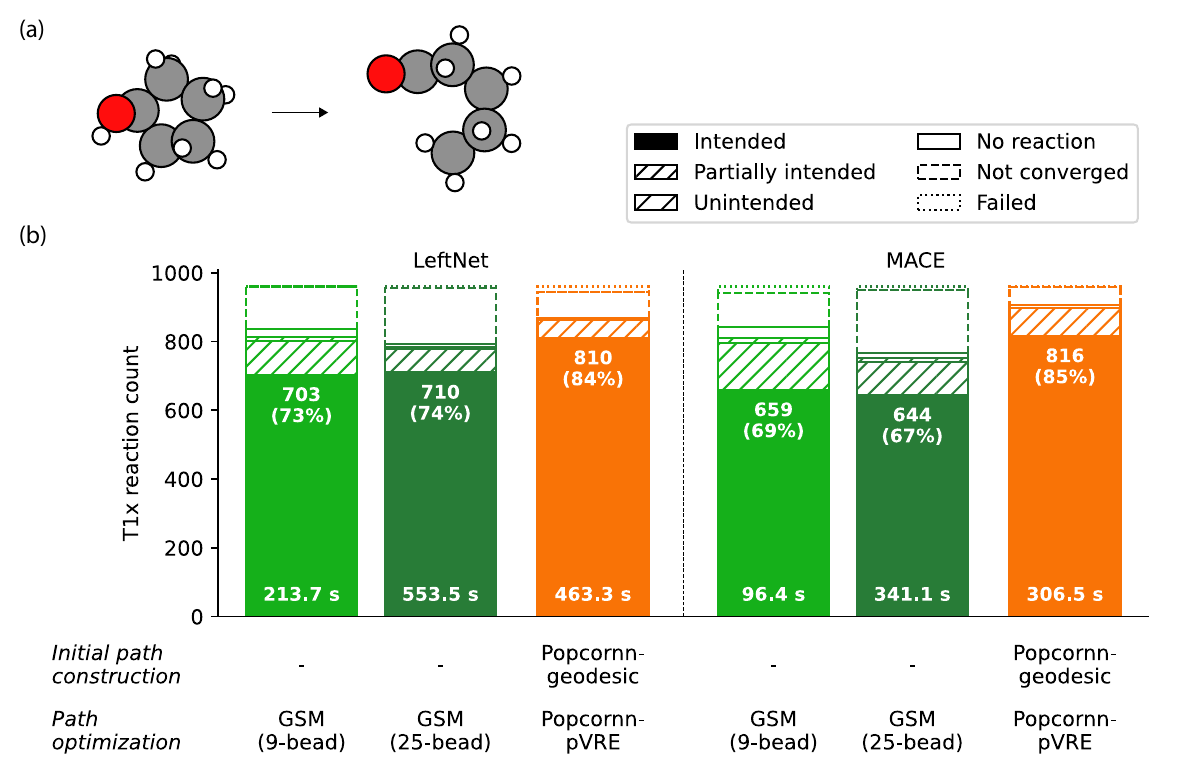}
    \caption{\textit{\ppcrn{} vs. GSM on organic reactions using LeftNet and MACE trained or fine-tuned on Transition-1x by Zhao et al \cite{Zhao2025}.} (a) An example of the test reactions evaluated here that are in T1x but not used in training. (b) \ppcrn{} improves recovery of ``Intended'' TSs for test reactions by 10\% and 16\% for LeftNet and MACE, respectively, and across varying numbers of beads (9 and 25). The number at the bottom of each bar is the average time, in seconds, for the TS optimization to converge.}
    \label{fig:bars_t1x}
\end{figure}

The second experiment compares \ppcrn{} and GSM performance on organic reactions using two popular MLIPs: LeftNet~\cite{Du2023} and MACE~\cite{Nigam2022,Kovcs2023}. Each MLIP was trained or fine-tuned on T1x by Zhao et al.~\cite{Zhao2025}, and we follow the training/validation split of T1x used by Zhao et al. for the set of test reactions. 
Rather than confirming the presence of a TS corresponding to each reaction on each separate MLIP PES, they confirmed the presence of the corresponding TS on the DFT PES used for MLIP training, yielding 960 DFT-validated test reactions.
Therefore, it is possible that a failure to identify an ``Intended'' here may be due to a failure of the double-ended TS search procedure or because the TS does not exist on the MLIP PES.
Zhao et al.~\cite{Zhao2025} showed that energy and force mean absolute errors were best for LeftNet and slightly worse for MACE. 

\ppcrn{} yields superior TS recovery over GSM across both MLIPs (Fig.~\ref{fig:bars_t1x}b).
\ppcrn{} achieves 84\% and 85\% ``Intended'' TSs compared with 74\% and 69\% for GSM with LeftNet and MACE, respectively.
While the absolute success of GSM falls from LeftNet to MACE, corresponding to decreasing accuracy of the MLIP, \ppcrn's success rate remains essentially unchanged.
This demonstrates how \ppcrn's numerical integration increases the robustness of the optimized path and TS by disproportionately improving upon less accurate MLIPs.
We further observe that the GSM ``Intended" rate is weakly affected by the number of beads, although the convergence time increases faster with the number of beads for MACE than it does for LeftNet.
\ppcrn{} is faster than 25-bead GSM, but roughly 2 to 3 times slower than 9-bead GSM.

%% file: 03_discussion.tex
\label{sec:discussion}

\ppcrn{} is a robust double-ended TS search framework explicitly designed to leverage the distinct advantages of MLIPs. 
MLIP efficiency and differentiability via simple backpropagation allow us to address fundamental weaknesses in previous approaches by employing a continuous path representation coupled with adaptive numerical integration of a path objective that directly targets the TS rather than the entire MEP. 
Further, the batched parallelization of our adaptive integrator allows \ppcrn{} to take full advantage of modern GPU hardware.
In contrast to typical small-scale comparisons of path optimization methods, we demonstrate \ppcrn's advantages via high-throughput evaluations with thousands of reactions at various levels of difficulty. While we evaluate robustness and computational runtime, we consider robustness to be the dominant factor that should govern method selection in the MLIP context.
This is because probing the PES will become extremely fast/cheap with future ML hardware and software advancements.
\ppcrn{} consistently outperforms NEB and GSM across multiple MLIPs and reaction sets while simultaneously reducing the number of hyperparameters that a user must choose, making it well-suited for high-throughput double-ended TS search. 

We attribute much of \ppcrn's success to the symbiotic combination of a continuous path representation \rxn{} and an adaptive numerical integrator.
The importance of incorporating careful numerical considerations within scientific machine learning methods has been highlighted previously~\cite{queiruga2020continuousindepthneuralnetworks,krishnapriyan2021characterizingpossiblefailuremodes,krishnapriyan2023learningcontinuousmodelscontinuous,sakarvadia2026falsepromisezeroshotsuperresolution, benitez2026neuralequilibrialongtermprediction, hegazy2026recencybiasedcausalattention}.
Although properly integrating Eq.~\ref{eq:loss_path_integral} requires a continuous path, the continuous path alone is not sufficient.
Earlier work by Ramakrishnan et al.~\cite{Ramakrishnan2025} employ neural network paths with fixed and discrete integral approximation schemes that are still vulnerable to missing essential PES structure along the path.
\ppcrn, instead, leverages the continuous nature of the path to enable adaptive numerical integration.
We therefore stress the importance of simultaneously having a continuous path \rxn{} and an accurate numerical integration scheme built into the path optimization.


\ppcrn{} is a highly flexible framework built to adapt to and grow with advancing machine learning, chemistry, and optimization techniques.
Specifically, by modularizing the path representation, PES, path integrator, and path objective, a user can swap any functionality for their own method. Various MLIPs, including NewtonNet~\cite{Haghighatlari2022, Yuan2024}, MACE~\cite{Nigam2022,Kovcs2023}, ANI-1x~\cite{Smith2017,Smith2018}, LeftNet~\cite{Du2023}, Orb~\cite{Neumann2024}, and UMA~\cite{Wood2025}, are already supported, and new models can be easily implemented.
While we used vanilla neural networks for the path representation in this work, spline representations are also available, and more sophisticated architectures are possible. 
Similarly, the objective integrator and the objective function can be replaced.
The modular architecture empowers chemists to systematically investigate how path representation, objective, optimization, integration, and PES calculations interact; an open methodological question that has thus far been difficult to address with monolithic implementations.
We envision the \ppcrn{} framework as a powerful software framework for subsequent path optimization methodology development as well as for establishing additional diverse reaction evaluations.

While the \ppcrn{} framework and the test systems used here are all molecular and non-periodic, we are currently generalizing \ppcrn{} to periodic systems, allowing it to be employed for double-ended transition state optimization for catalysis and material systems.
For such periodic non-molecular systems, future work will determine if the geodesic loss and the pairwise potential from~Zhu et al.~\cite{Zhu2019} (designed for small molecules) will be sufficient, or if a new interpolation loss (e.g., a continuous implementation of IDPP) will be needed.

Reaction networks are broadly constructed in one of two ways~\cite{Wen2023}.
Exploration-based construction grows the network outward from the reactants with a single-ended TS search, discovering rather than presupposing products~\cite{Unsleber2022,Han2026nexus}.
To avoid combinatorial blowup in the number of reactions, one must constrain which transformations are sampled and which species are expanded as the network deepens.
Enumeration-based construction instead bounds the chemical space up front by enumerating candidate species and reactant--product pairs~\cite{Zhao2021,Blau2021,Barter2023}.
Each pair then requires a double-ended TS search, which imposes no restriction on the bond-change pattern connecting the endpoints.
\ppcrn{} is designed for this second setting.
It is also complementary to generative TS models~\cite{Duan2023,Kim2024,Duan2025,Galustian2025,Han2026nexus}, which can rapidly propose accurate TSs within the chemistry represented in their training data but require curated sets of converged TSs to train.
Since \ppcrn{} is a tool that consumes MLIPs, its scope is only set by the underlying pre-trained MLIP, rather than any one particular dataset coverage.
For example, \ppcrn{} already works with foundation MLIPs trained on OMol25~\cite{Levine2025,Wood2025} already span transition metals and diverse charge and spin states, chemistry for which reactive TS data remain scarce.

Finally, as MLIPs continue to improve in speed, accuracy, and generality \cite{yuan2025foundationmodelsatomisticsimulation}, we believe that robust transition state search via \ppcrn{} and subsequent improvements atop our software framework will enable the construction of massive chemical reaction networks, allowing us to understand and eventually design complex reactive processes in electrolytes, catalysis, polymers, and beyond.

%% file: 04_methods.tex

\subsection{Interfacing with \ppcrn}

Here, we describe \ppcrn{} in greater detail, walk through important design choices, and outline key steps in applying \ppcrn{}.
This section generally follows the \ppcrn{} workflow and highlights key components that a user may wish to adapt to their specific problem.
Supplementary Note~\ref{app:algorithms} gives pseudocode for each step, and Algorithm~\ref{alg:popcornn} summarizes a complete run.

\subsubsection{Input Endpoints into \ppcrn}


The \ppcrn{} package takes optimized endpoints as inputs.
\ppcrn{} itself does not optimize the endpoints and expects the user to have done so with their preferred method.
In this work, the endpoints are first relaxed with BFGS on the MLIP PES.
\ppcrn{} fixes the input endpoints in Cartesian coordinates, thus requiring the endpoints to be translationally and rotationally aligned.
See Supplementary Note~\ref{app:path} for more discussion.

\subsubsection{The Path Model\label{sec:path_model}}


A single 2-layer perceptron is used for the path representation, following Eq. \ref{eq:path}.
Concretely, the neural network $\mathbf{f}_\theta$ is $Linear(1, 128) \to GELU() \to Linear(128, 3N)$.
The GELU activation is chosen because of its smoothness \cite{hendrycks2023gaussianerrorlinearunits}.
Supplementary Note~\ref{app:path} provides more details on the path representation, and Algorithm~\ref{alg:path_eval} (Supplementary Note~\ref{app:alg_path}) shows how positions, velocities, energies and forces are evaluated along the path.

\subsubsection{Path Initialization\label{sec:path_interpolation}}


\ppcrn{} implements a continuous path version of the geodesic approach for initial path construction \cite{Zhu2019}.
The geodesic approach uses the pairwise repulsive potential
\begin{equation}
    U_\text{geo} = \sum_{i<j} \exp \left[ -\alpha \frac{r_{ij} - r^{(\text{eq})}_{ij}}{r^{(\text{eq})}_{ij}} \right] + \beta \frac{r^{(\text{eq})}_{ij}}{r_{ij}} ,
    \label{eq:pairpotential}
\end{equation}
where $r_{ij}$ and $r^{(\text{eq})}_{ij}$ are the current and approximate equilibrium distances, respectively, between atoms $i$ and $j$ within a cutoff distance of 3 \r{A}, while $\alpha=1.7$ and $\beta=0.01$ scale the contributions of each energy term, as originally proposed by Zhu et al.~\cite{Zhu2019}.
In the original geodesic approach, discrete points were optimized.
In \ppcrn, we apply the same loss to a continuous path
\begin{equation}
    \mathcal{L}_\text{geo} 
    = \int_{0}^{1} \sqrt{ \sum_{q \in \{(i,j)|i < j\}} \left| \mathbf{F}_q(\mathbf{x}_\theta(t)) \cdot \mathbf{v}_\theta(t) \right|^2 } \, dt
    = \int_{0}^{1} \left\| \mathbf{F}^\text{(pairs)}(\mathbf{x}_\theta(t)) \cdot \mathbf{v}_\theta(t) \right\| \, dt ,
    \label{eq:geodesic}
\end{equation}
where $\|\cdot\|$ is the norm of a $(N(N-1)/2)$-dimensional vector across unique atomic pairs, and $\mathbf{F}^\text{(pairs)} \in \mathbb{R}^{3N \times \frac{N(N-1)}{2}}$ is a matrix by stacking pairwise forces $\mathbf{F}_q \in \mathbb{R}^{3N}$.
See Supplementary Note~\ref{app:geo_loss} for more detail.
This initial path construction is the first optimization stage of Algorithm~\ref{alg:popcornn} (Supplementary Note~\ref{app:alg_overview}).

\subsubsection{Interfacing with MLIPs}


We have interfaced \ppcrn{} with various MLIPs, including NewtonNet~\cite{Haghighatlari2022, Yuan2024}, MACE~\cite{Nigam2022,Kovcs2023}, ANI-1x~\cite{Smith2017,Smith2018}, LeftNet~\cite{Du2023}, Orb~\cite{Neumann2024}, and UMA~\cite{Wood2025}. The model weights are acquired from Yuan et al.~\cite{Yuan2024}, Zhao et al.~\cite{Zhao2025}, and Wood et al.~\cite{Wood2025}.
The \ppcrn{} repository includes a template class and necessary functions to interface with future MLIPs.
The \ppcrn{} code repository~\cite{hegazy2026popcornn} provides more details, and Supplementary Note~\ref{app:alg_code} lists where each step is implemented.

\subsubsection{TS Determination}
\label{sec:TS_search}

When searching for the TS along \rxn, we found that both minimizing the PES gradient and maximizing energy led to the most robust method.
The TS is the highest energy point along \rxn, but simply selecting the highest energy point among the evaluated \rxn{} points may not be sufficiently precise if the adaptive integration evaluations are too sparse.
Instead, we find points between the evaluated path points at which $dE/dt = 0$ (calculated using Eq.~\ref{eq:energy_derivative}) since this is more stable and has higher precision than a finite difference comparison between energy calculations.
First, we determine where $dE/dt$ changes sign within the cached values from the latest \ppcrn{} optimization evaluation.
This can result in multiple TS candidates.
Second, we find the TS time of each pair by linearly interpolating between the cached time points and solve for $t$ that satisfies $dE(t)/dt=0$.
Finally, we evaluate the PES energy of all the TS time candidates and determine $t^*$ as the highest energy~point; this discards sign changes at energy minima.
Algorithm~\ref{alg:ts_search} (Supplementary Note~\ref{app:alg_ts}) gives the full procedure.

\subsection{\ppcrn{} Path Optimization}

\subsubsection{Parallel Evaluations and Memory Management}

The loss integration is handled by \pdq~\cite{Hegazy2025}: a parallelized adaptive quadrature numerical integrator for path integrals with backpropagation capabilities.
The adaptive numerical integration is necessary to properly evaluate Eq.~\ref{eq:loss_path_integral}.
\pdq{} leverages the knowledge of the path to parallelize the integration, improving speed by orders of magnitude.
\pdq{} also uses an adaptive memory management scheme to fully utilize GPU parallelism.
We apply the absolute and relative tolerances of $0$ and $1$, respectively, for the numerical integration.
Hegazy et al.~\cite{Hegazy2025} and Supplementary Note~\ref{app:pdq_parallelism} provide more detailed discussions.
Algorithms~\ref{alg:path_gradient} and~\ref{alg:quadrature} (Supplementary Notes~\ref{app:alg_gradient} and~\ref{app:alg_quadrature}) give the gradient evaluation and the adaptive quadrature.

\subsubsection{Optimization Hyperparameters}

We use the Adam optimizer with a learning rate of $10^{-3}$ for both initial path construction atop the initialization potential and final optimization atop the MLIP potential.
Each of these optimizations used its own optimizer; that is, the optimizer was not shared between them.
We run each optimization for up to 1000 steps, or stop as it reaches the gradient convergence of $10^{-2}$ for initial path construction and $10^{-1}$ for path optimization (corresponding to TS forces of 0.05 eV/\AA).
Supplementary Note~\ref{app:convergence} provides more detailed discussions, and Algorithm~\ref{alg:popcornn} shows where each of these settings enters.

\subsection{TS Optimization}

The TS acquired from \ppcrn{} or other reference methods are optimized to a first-order saddle point on the MLIP using Sella \cite{Hermes2022} with the full Hessian at every step \cite{Yuan2024} with a convergence threshold of 0.01 eV/\AA{} for up to 1000 steps.
For retrieved TSs, forward and backwards IRC calculations are run using SellaIRC~\cite{Hermes2022} with a convergence threshold of 0.01 eV/\AA{} for up to another 1000 steps. For organic reactions, the endpoints are compared using connectivity graphs created by Open Babel \cite{OBoyle2011}. 

\subsection{Reference Methods\label{sec:reference_methods}}

For a fair comparison, we provide the same inputs to each method and apply the same TS optimization to all outputs. Specifically, we give each method the same optimized endpoints. The output TSs of each method are processed in exactly the same way.

We use ``improved tangent'' NEB \cite{Jonsson1998} with 10, 20, and 100 beads (k=5.0) and GSM \cite{Zimmerman2013} with 9 and 25 beads and a convergence threshold of 0.05 eV/\AA. NEBs are optimized with the MDMin optimizer (dt=0.02, maxstep=0.1) in ASE for 2000 steps in total, where the first 1000 steps are without the climbing image and the last 1000 steps are with the climbing image, with a convergence threshold of 0.05 eV/\AA{} \cite{Henkelman2000}. All other hyperparameters are in ASE 3.28.0 and pyGSM defaults.

%% file: data_availability.tex
All reaction data and model weights used in this study are publicly available.
The reactions come from the Transition1x dataset~\cite{Schreiner2022,Schreiner2022data} and the Grambow--Green dataset~\cite{Grambow2020,Grambow2020data}; the two test sets are subsets of these datasets, selected as described in Results.
The NewtonNet weights trained on Transition1x by Yuan et al.~\cite{Yuan2024} are available at \url{https://github.com/THGLab/NewtonNet/releases/download/pretrained/newtonnet_t1x.zip}.
The LeftNet and MACE weights trained or fine-tuned on Transition1x by Zhao et al.~\cite{Zhao2025} are available at \url{https://huggingface.co/yhong55/ReactBench}, with usage instructions at \url{https://github.com/deepprinciple/ReactBench}.

%% file: code_availability.tex
\ppcrn{} is available at \url{https://github.com/khegazy/popcornn}~\cite{hegazy2026popcornn}, and the GPU-parallelized adaptive numerical integrator \pdq{} is available at \url{https://github.com/khegazy/padaquad}~\cite{Hegazy2025}.
Both are released under the Creative Commons Attribution 4.0 International licence.
The MLIP implementations are available from their public repositories: NewtonNet~\cite{Haghighatlari2022} at \url{https://github.com/THGLab/NewtonNet}, LeftNet~\cite{Du2023} at \url{https://github.com/yuanqidu/LEFTNet}, and MACE~\cite{Nigam2022,Kovcs2023} at \url{https://github.com/ACEsuit/mace}.

%% file: acknowledgements.tex
We thank Graeme Henkelman and Ilgar Baghishov for helpful discussions about NEB hyperparameters. We thank Paul Zimmerman for helpful discussions about GSM hyperparameters. This work used computational resources provided by the National Energy Research Scientific Computing Center (NERSC), a U.S. Department of Energy Office of Science User Facility operated under Contract DE-AC02-05CH11231, and the Lawrencium computational cluster resource provided by the IT Division at the Lawrence Berkeley National Laboratory (Supported by the Director, Office of Science, Office of Basic Energy Sciences, of the U.S. Department of Energy under Contract No. DE-AC02-05CH11231).

%% file: funding.tex
All authors acknowledge support from the Laboratory Directed Research and Development program of Lawrence Berkeley National Laboratory (LBNL), supported by the Office of Science, Office of Basic Energy Sciences (BES), of the U.S. Department of Energy (DOE) under Contract No.\ DE-AC02-05CH11231.
S.M.B. further acknowledges support from the Center for High Precision Patterning Science (CHiPPS), an Energy Frontier Research Center funded by the U.S. Department of Energy, Office of Science, Basic Energy Sciences at Lawrence Berkeley National Laboratory under Contract No. DE-AC02-34205CH11231.
C.-Y.E.Y and T.H.G. further acknowledge support from the CPIMS program and funded by the U.S. Department of Energy, Office of Science, Basic Energy Sciences at Lawrence Berkeley National Laboratory under Contract No. DE-AC02-05CH11231.
D.M.\ further acknowledges support from the Scientific Discovery through Advanced Computing (SciDAC) program funded by the U.S. Department of Energy, Office of Science, Office of Advanced Scientific Computing Research at Lawrence Berkeley National Laboratory under Contract No.\ DE-AC02-05CH11231.
M.W.M. acknowledges DARPA, NSF, the DOE Competitive Portfolios grant, and the DOE SciGPT grant.

%% file: S01_path_representation.tex
\label{app:path}

We parameterize our reaction path as a learned deviation from a linear interpolation between the endpoints.
This parameterization was introduced in Eq.~\ref{eq:path} and repeated here for clarity
\begin{equation*}
    \rxnmm = \underbrace{(1 - t) \, \xrmm + t \, \xpmm}_\text{linear prior} + \underbrace{t \, (1-t) \, \fnnmm}_\text{perturbation} .
\end{equation*}
This parameterization anchors both the endpoints and the path.
Our path model (\fnn) is a simple two-layer MLP with GELU activations.
While optimizing the \ppcrn{} path representation \rxn, we observed that activation functions that are not smooth (e.g., ReLU and SELU) underperform smooth activations like GELU.
We use the Cartesian coordinate system to represent both the input and output structures.

\subsection{Ablation Study of the Endpoint Enforcement}

A double-ended TS finding method must fix the two endpoints, as in Fig.~\ref{fig:model}a. 
This is trivially accomplished for discrete approaches like NEB, simply by not applying any updates to the endpoint beads. To impose the constraint in the continuous case, however, is not as straightforward. Our linear path formulation (Eq.~\ref{eq:path}) enforces the endpoint structures as hard constraints ($x_\theta(0) = x_R$ and $x_\theta(1) = x_P$) through a quadratic envelope. To demonstrate the flexibility and robustness of our methodology, we ablate this envelope by removing it and replacing it with other envelopes.

\begin{figure}[tb!]
    \centering
    \includegraphics[width=\linewidth]{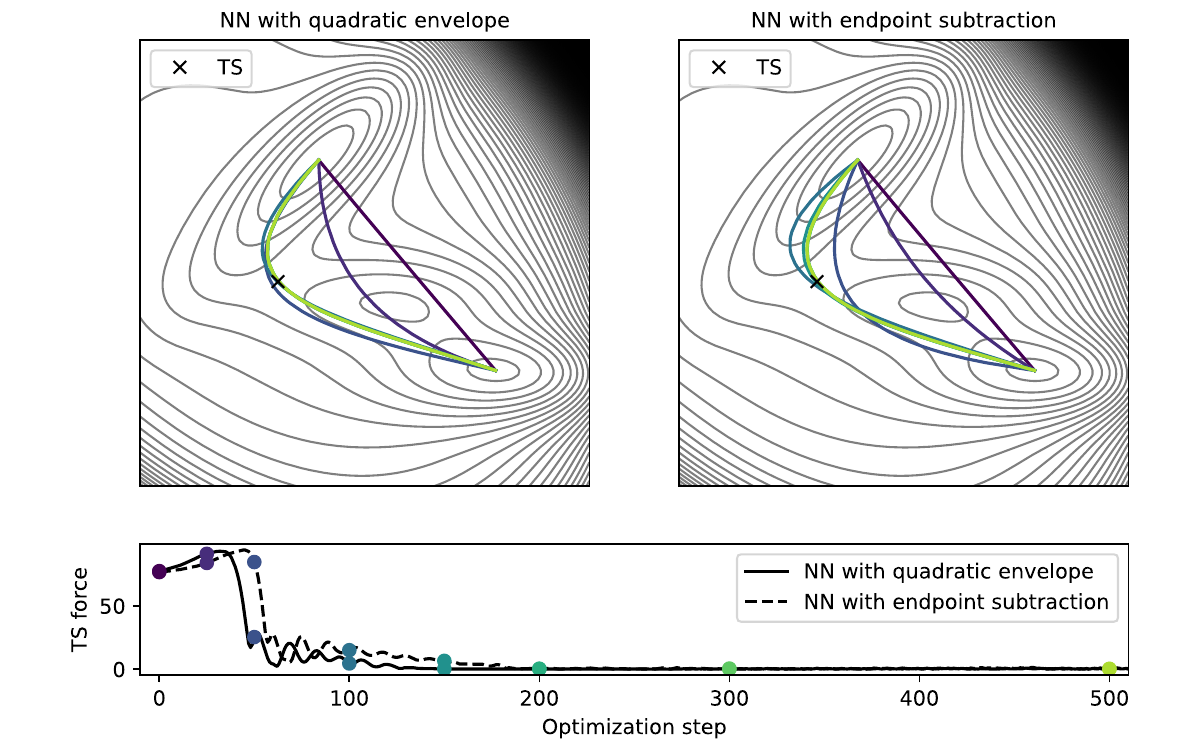}
    \caption{\textit{Comparison between path constraint implementations.} The path with a quadratic envelope $t \, (1-t)$ in Eq. \ref{eq:path} (top left) can be optimized more stably than the path with endpoint subtraction in Eq. \ref{eq:path_sub} (top right). The bottom row compares the TS force during dynamics. The colored lines and circles are corresponding measurements at the same time.}
    \label{fig:subtraction}
\end{figure}

We remove the endpoint-enforcing envelope and enforce the endpoints by subtracting the neural network values.
The resulting path formulation
\begin{equation}
    \rxnmm = \underbrace{(1 - t) \, \xrmm + t \, \xpmm}_\text{linear prior} + \underbrace{\fnnmm - (1-t) \, \mathbf{f}_\theta(0) - t \, \mathbf{f}_\theta(1)}_\text{perturbation} ,
    \label{eq:path_sub}
\end{equation}
enforces $x_\theta(0) = x_R$ and $x_\theta(1) = x_P$. This hard constraint enforcement leads to slight optimization instabilities compared to Eq.~\ref{eq:path}, shown in Fig.~\ref{fig:subtraction}.
We presume these instabilities are due to the interdependence between $\mathbf{f}_\theta(t)$, $\mathbf{f}_\theta(0)$, and $\mathbf{f}_\theta(1)$.

\begin{figure}[p]
    \centering
    \includegraphics[width=\linewidth]{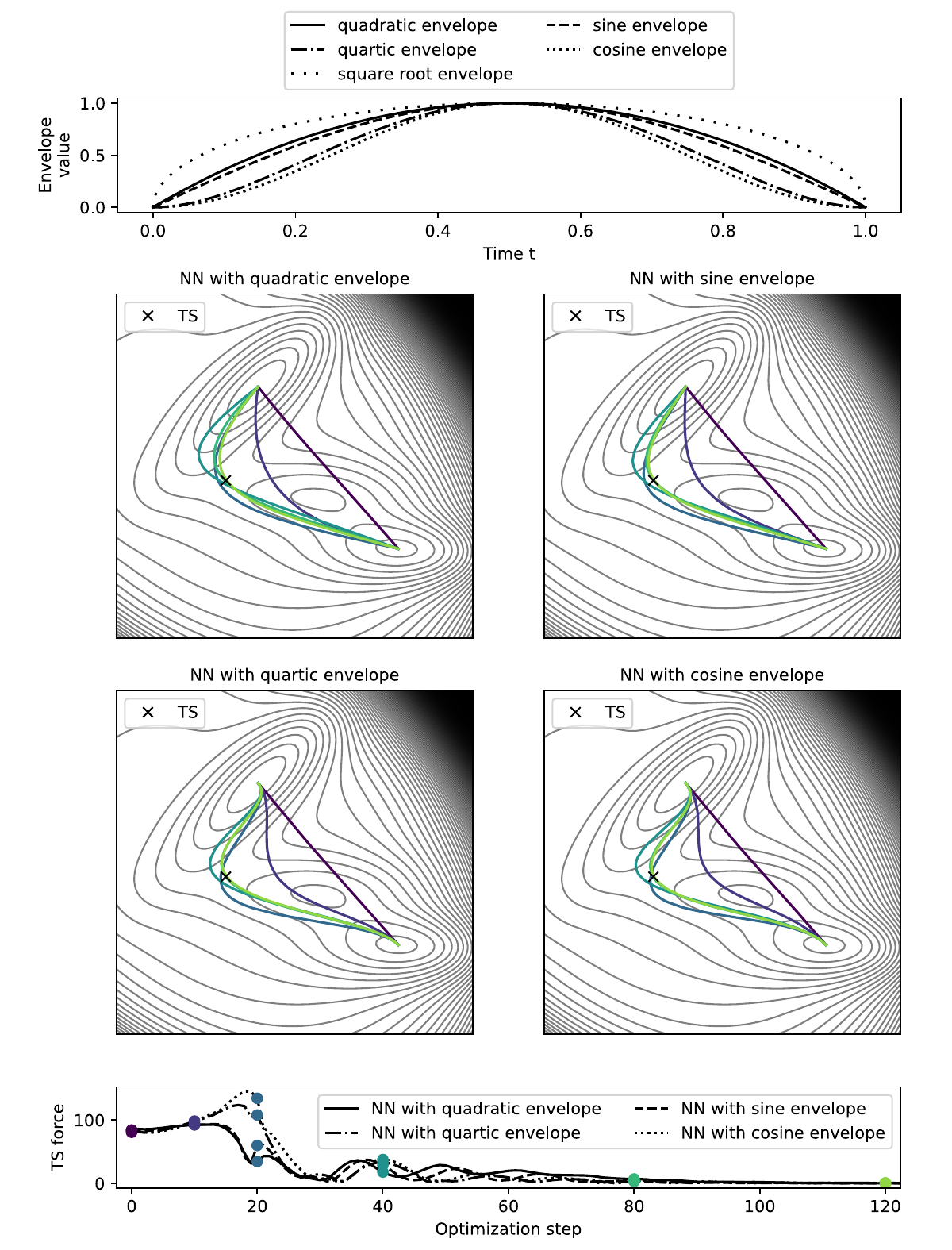}
    \caption{\textit{Comparison between path envelope forms.} We compare multiple path envelope ablations (top panel) and their optimization dynamics (middle and lower panels). For rigorous comparisons, we scale all envelope functions to have maxima of 1 at $t=0.5$, roughly equivalent to scaling the learning rate. The colored lines and circles are corresponding measurements at the same time.}
    \label{fig:envelope}
\end{figure}

We ablate the envelope with functions of varying polynomial degrees and function types.
Our ablation envelopes,
\begin{align}
    \text{Quadratic Envelope:}&\quad t\,(1-t) \label{eq:quad_env}\\
    \text{Normalized Quadratic Envelope:}&\quad 4\,t\,(1-t) \\
    \text{Quartic Envelope:}&\quad t^2 \, (1-t)^2 \\
    \text{Sine Envelope:}&\quad \sin(\pi t) \\
    \text{Cosine Envelope:}&\quad 1-\cos(2\pi t) ,
\end{align}
illustrated in the top panel of Fig.~\ref{fig:envelope}, primarily differ at the endpoints. We ablate the quadratic term with the normalized quadratic because for the original quadratic envelope $x_\theta(0.5) = 0.25$ and for the normalized $x_\theta(0.5) = 1$, similar to the other envelopes. When testing the path optimization on the Muller-Brown potential, all the envelopes behave similarly (Fig.~\ref{fig:envelope}). When comparing the quadratic and sine envelopes with the quartic and cosine envelopes, the primary difference is at the endpoints. For the quartic and cosine envelopes, the path is artificially straightened. This may become a problem in some real-world optimizations where either the product or reactant state is fairly unstable. This is because the quadratic and sine envelopes have finite velocities at $t=0$ and $t=1$, whereas the quartic and cosine envelopes have zero velocity. Therefore, we would expect some envelopes to produce a path that cannot be optimized, like a square root $(\sqrt{t \, (1-t)})$, which has infinite velocities at the endpoints.

\subsection{Linear Prior Ablation Study}

The linear parameterization helps keep the path from deviating to something unphysical. We refer to this as our linear prior since the linear interpolation biases the initial path towards physicality. This linear prior offers a rough direction of the atom movements, and the neural network is trained to correct the $3N$-dimensional movements as a function of the reaction coordinate $t$. This formalism is also adopted by other related work \cite{Han2024,Ramakrishnan2025,Nam2025}.

\begin{figure}[tp!]
    \centering
    \includegraphics[width=\linewidth]{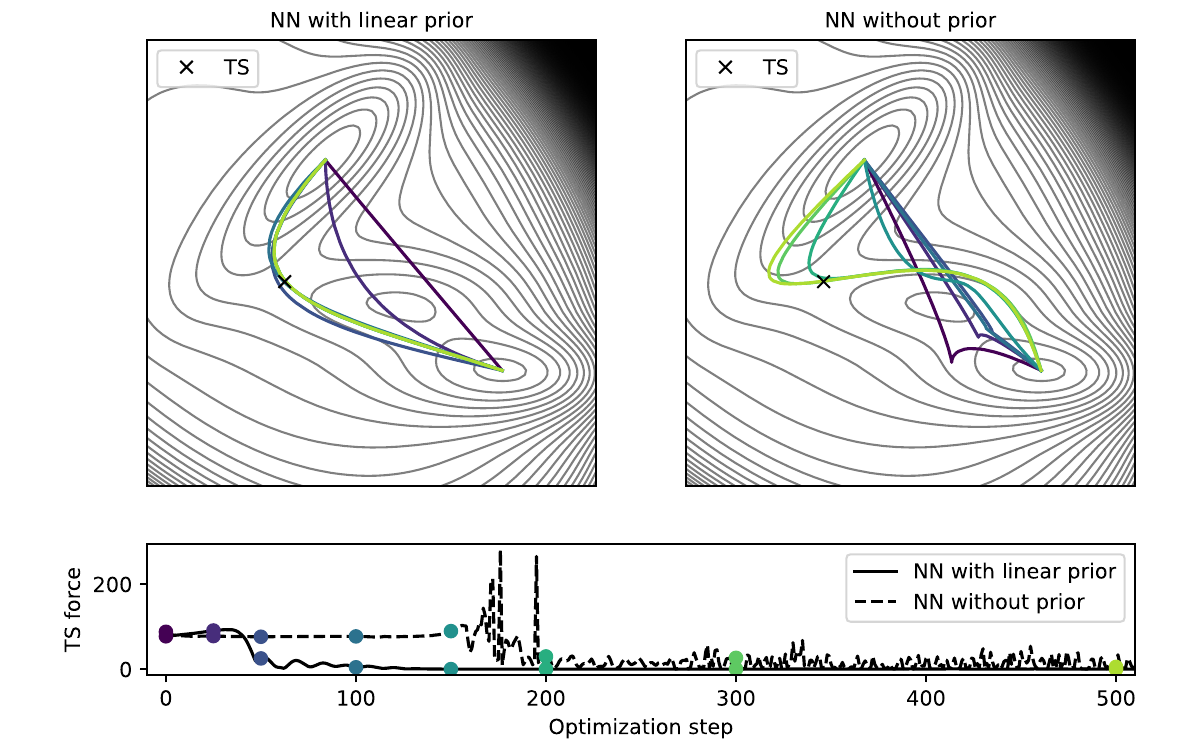}
    \caption{\textit{Comparison between path formalisms.} We show the path optimization atop the model Muller-Brown potential using \ppcrn{} with (top left) and without (top right) the linear prior in Eq.~\ref{eq:path}. The bottom panel shows the TS force during optimization. The colored lines and circles are corresponding measurements at the same time.}
    \label{fig:prior}
\end{figure}

To illustrate the linear prior's importance, we consider the bias-variance tradeoff by removing it.
The randomized path,
\begin{equation}
    \rxnmm = [ 1 - 4 t \, (1-t) ] [ (1 - t) \, \xrmm + t \, \xpmm ] + 4 t \, (1-t) \, \mathbf{f}_\theta(t) ,
    \label{eq:path_random}
\end{equation}
allows the middle point to be randomly initialized: $(\mathbf{x}_\theta(0.5)=\mathbf{f}_\theta(0.5))$.
This random initialization removes the linear bias, or anchoring effect, of the linear prior.
However, this random initialization is likely to initialize $\mathbf{x}_\theta(0.5)=\mathbf{f}_\theta(0.5)$ in a challenging region of the optimization landscape: e.g., at an unphysical structure, a very unlikely structure, or a structure that requires highly unlikely bond breaking and formation.
To validate this intuition, we used the random initialization of Eq.~\ref{eq:path_random} on the Muller-Brown potential, where we observe challenging and chaotic optimization (Fig.~\ref{fig:prior}).
Ultimately, the sporadic optimization of Eq.~\ref{eq:path_random} results in a more convoluted path.
Given these challenges with this simple model potential, we expect these issues to be exacerbated when using an MLIP.

\subsection{Hidden Units as a Time Basis}
\label{app:path_architecture}

Since the action of the \rxn{} neural network is a linear combination of the nonlinear hidden units ($\mathbf{h}(t)$), \rxn{} can be interpreted as a weighted sum of learned reaction bases.
Here we provide more detail on the path model with our two-layer neural network.

A two-layer neural network, with activations $\phi$, has 4 learnable objects: a weight matrix $\mathbf{W}$ and a bias $\mathbf{b}$ for each layer.
To improve stability for the path neural network, the scalar reaction coordinate $t$ is first rescaled to
\begin{equation}
    \tilde t = 2t - 1.
\end{equation}
Passing $\tilde{t}$ into the first layer of the path network gives us the first set of hidden units
\begin{equation}
    z_j(\tilde{t})
    =
    W^{(1)}_j \tilde t + b^{(1)}_j,
    \qquad
    h_j(\tilde{t})
    =
    \phi(z_j(\tilde{t})),
\end{equation}
where $j=1,\ldots,w$ indexes the hidden units and $\phi$ is a smooth activation function. The output of the neural network is then
\begin{equation}
    f_{\theta,i}(\tilde{t})
    =
    b^{(2)}_i
    +
    \sum_{j=1}^{w}
    W^{(2)}_{ij} h_j(\tilde{t}),
\end{equation}
for Cartesian coordinate $i$.

Thus, the hidden units from the first layer ($\mathbf{h}(\tilde{t})$) form a trainable basis set of the reaction coordinate. 
When using a non-smooth activation function (e.g., ReLU or SELU), the basis itself will not be smooth.
This additionally makes the atomic velocities and the loss integrand discontinuous, ultimately making the integration, optimization, and convergence more challenging.
In our setting with GELU activation, each $h_j(\tilde{t})$ is a smooth feature over time, and the output layer constructs each Cartesian coordinate of the path as a linear combination of these features.

To enforce the \rxn{} endpoints as a hard constraint, we use the quadratic envelope $q(t)$ (Eq.~\ref{eq:quad_env}).
Thus, the full \rxn{} is
\begin{align}
    q(t) &= t(1 - t) \\
    x_{\theta,i}(t)
    &=
    (1-t)x_{R,i}
    +
    t x_{P,i}
    +
    q(t)
    \left[
    b^{(2)}_i
    +
    \sum_{j=1}^{w}
    W^{(2)}_{ij} h_j(\tilde{t})
    \right].
\end{align}
This is analogous to representing the path with a set of smooth basis functions in $t$, except that the basis functions are learned jointly with
the path.
Increasing the hidden width increases the number of available time features, while the output layer selects their amplitudes independently for
each Cartesian coordinate.

\subsection{Internal Versus Cartesian Coordinates}
Our path operates in Cartesian coordinates instead of the internal coordinates.
Although empirical evidence has shown that internal coordinates are important for geometry optimization \cite{Hermes2022}, transforming from internal coordinates to Cartesian is non-unique and generally difficult. This difficulty is in part because internal coordinates can have singularities, and optimal coordinates generally change throughout the reaction.

One of the primary benefits of internal coordinates is the grouped movement of atoms by one or a few coordinates.
This feature can be captured by the neural network path, which can learn the optimal time-dependent internal representation.
Thus, the neural network path effectively learns the time-dependent internal coordinates and their transformation.

The input molecules must be properly positioned before being passed to \ppcrn. This includes handling the translational and rotational degrees of freedom. This is a non-trivial task that is best handled by specialty external packages~\cite{Simm2017,Unsleber2022}.

\subsection{Complexity of the Path Model}
The neural network \fnn{} is much smaller than the MLIP, serving only as a flexible representation of the path, not as a sophisticated regressor. Here, we use a simple fully-connected neural network instead of any invariant or equivariant architecture. Such inductive biases are commonplace in many MLIPs~\cite{Nam2025, Nigam2022}. We found that many of these inductive biases can be detrimental to our path optimization. Moreover, allowing the neural network to learn optimal coordinate systems and transformations removes the need for such inductive biases.

The locality assumption is a very common inductive bias found in most all MLIPs. Here, the representation of each atom is updated by information of atoms within some radius. This, however, can be detrimental for \ppcrn, which can optimize the reaction between two separated molecules. Imagine the simplest possible reaction \ch{H + H -> H2} where the reactant atoms are separated much further than a typical radius used in the locality assumption. This path, therefore, is generally not local. Although some of the non-locality is captured by the linear prior, it is likely that this inductive bias would hamper the optimization.

Another common and important inductive bias is rotation and translation invariance. However, the linear prior breaks the individual translational and rotational symmetries of endpoints. Given the importance of the linear prior in this work, we consider that incorporating invariant or equivariant architectures for the perturbation neural network will lead to complications.

\subsection{Path Jacobian}

The path Jacobian quantifies the sensitivity of each atomic coordinate to the learnable weights, and thus it plays an important role in the optimization.
Since the linear prior is independent of $\theta$ (Eq.~\ref{eq:path}), the path Jacobian is
\begin{equation}
    J_\theta(t)
    =
    \frac{\partial \mathbf{x}_\theta(t)}{\partial \theta}
    =
    q(t)
    \frac{\partial \mathbf{f}_\theta(\tilde t)}{\partial \theta}.
    \label{eq:jacobian}
\end{equation}
The quadratic envelope (Eq.~\ref{eq:quad_env}) scales the sensitivity of the path to the neural-network parameters, and removes $\mathbf{f}_\theta(\tilde t)$'s contribution from the endpoints.

For the two-layer MLP used here,
\begin{equation}
    \mathbf{z}(\tilde{t})
    =
    \mathbf{W}^{(1)} \tilde t + \mathbf{b}^{(1)},
    \qquad
    \mathbf{h}(\tilde{t})
    =
    \phi(\mathbf{z}(\tilde{t})),
\end{equation}
\begin{equation}
    \mathbf{f}_\theta(\tilde t)
    =
    \mathbf{W}^{(2)} \mathbf{h}(\tilde{t})
    +
    \mathbf{b}^{(2)},
\end{equation}
where $\phi$ is the GELU activation. If $i$ indexes Cartesian output
coordinates and $j$ indexes hidden units, the nonzero Jacobian blocks are
\begin{align}
    \frac{\partial x_{\theta,i}(t)}
    {\partial W^{(2)}_{ij}}
    &=
    q(t) h_j(\tilde{t}), \label{eq:rxn_jac_W2}
    \\
    \frac{\partial x_{\theta,i}(t)}
    {\partial b^{(2)}_i}
    &=
    q(t),
    \\
    \frac{\partial x_{\theta,i}(t)}
    {\partial W^{(1)}_{j}}
    &=
    q(t) W^{(2)}_{ij} \phi'(z_j(\tilde{t})) \tilde t \label{eq:rxn_jac_second_order_W2}
    \\
    \frac{\partial x_{\theta,i}(t)}
    {\partial b^{(1)}_{j}}
    &=
    q(t) W^{(2)}_{ij} \phi'(z_j(\tilde{t})).
\end{align}

The final-layer weight block (Eq.~\ref{eq:rxn_jac_W2}) gives a simple estimate of the path Jacobian scale, because the scale of the weights ($W^{(2)}_{ij}$) is dependent on its input dimension $\mathcal{O}(1/\sqrt{w})$, while the GELU hidden units $h_j$ are $\mathcal{O}(1)$.
Generally, $w > 100$, making Eq.~\ref{eq:rxn_jac_W2} at least an order of magnitude larger than Eq.~\ref{eq:rxn_jac_second_order_W2}.
\begin{equation}
    J_\theta (t) \approx J_{\mathbf{W}^{(2)}}(t) = \frac{\partial x_{\theta,i}(t)}
    {\partial W^{(2)}_{ij}} .
\end{equation}
Taking the inner product of the Jacobian with itself results in a positive semidefinite matrix, allowing us to make a comparison of the singular values $\sigma$:
\begin{align}
    \sigma_i\left(J_{\theta}(t) J_{\theta}(t)^\top \right)
    &\gtrsim
    \sigma_i \left( J_{\mathbf{W}^{(2)}}(t) J_{\mathbf{W}^{(2)}}(t)^\top \right)
    =
    q(t)^2 \|\mathbf{h}(\tilde{t})\|_2^2 \\
    \sigma_{\min} \left( J_\theta(t) \right)
    &\gtrsim
    \sigma_{\min}\left(J_{\mathbf{W}^{(2)}}(t) \right)
    =
    q(t)\|\mathbf{h}(\tilde{t})\|_2 ,
\end{align}
where the equality holds because the inner product of the Jacobians is diagonal across Cartesian output coordinates.
At the path midpoint, $q(1/2)=1/4$, $\tilde t=0$, and $\mathbf{h}(1/2)=\phi(\mathbf{b}^{(1)})$.
The hidden pre-activations reduce to the first-layer biases:
\begin{equation}
    \sigma_{\min}(J_\theta(1/2))
    \gtrsim
    \frac{1}{4}
    \|\phi(\mathbf{b}^{(1)})\|_2 .
\end{equation}
Thus, at the reaction time midpoint, the scale of the smallest singular value is set by both the hidden width and the activation function.
For PyTorch's default initialization of a first-layer map from one scalar
input to $w$ hidden units, the biases satisfy approximately
$b^{(1)}_j \sim \mathcal{U}[-1,1]$. Thus,
\begin{equation}
    \mathbb{E}
    \left[
    \|\mathbf{h}(1/2)\|_2^2
    \right]
    =
    w
    \,
    \mathbb{E}_{b \sim \mathcal{U}[-1,1]}
    \left[
    \phi(b)^2
    \right],
\end{equation}
and the expected midpoint scale is
\begin{equation}
    \sigma_{\min}(J_\theta(1/2))
    \approx
    \frac{1}{4}
    \sqrt{
    w
    \,
    \mathbb{E}_{b \sim \mathcal{U}[-1,1]}
    \left[
    \phi(b)^2
    \right]
    } .
\end{equation}
For the GELU activation,
\begin{equation}
    \phi(b)
    =
    \frac{b}{2}
    \left[
    1 + \operatorname{erf}
    \left(
    \frac{b}{\sqrt{2}}
    \right)
    \right],
\end{equation}
which gives
\begin{equation}
    \mathbb{E}_{b \sim \mathcal{U}[-1,1]}
    \left[
    \phi(b)^2
    \right]
    =
    \frac{1}{2}
    \int_{-1}^{1}
    \phi(b)^2 \, db
    \approx
    0.1088 .
\end{equation}
Therefore, for the default width $w=128$,
\begin{equation}
    \sigma_{\min}(J_\theta(1/2))
    \approx
    \frac{1}{4}
    \sqrt{128 \times 0.1088}
    \approx
    0.93 .
    \label{eq:singular_value}
\end{equation}
Therefore, both the size of the number of hidden unit bases and the activation itself determine the Jacobian's scale.
As we will show in Supplementary Note~\ref{app:convergence}, this Jacobian plays a crucial role in the convergence criteria.

%% file: S02_loss.tex
\label{app:loss}

For double-ended TS searches, we use the assumption of simple TS theory that the path is simply a tool for finding the TS, and other points along the entire path (e.g., MEP, the gradient descent path from the TS, etc.) are not needed. Hence \ppcrn{} is designed to work with all losses that can be written as a path integral (Eq.~\ref{eq:loss_path_integral}).
In this work, we investigate three path integral losses: \vre{} in Eq.~\ref{eq:vre}; \pvre{} in Eq.~\ref{eq:pvre}; and \geo{} in Eq.~\ref{eq:geodesic}. We have demonstrated their behaviors in Fig.~\ref{fig:MB_vre_pvre}, and we provide further discussion on their physical interpretation below.

\subsection{The Variational Reaction Energy (VRE) Loss}

\begin{figure}[tb!]
    \centering
    \includegraphics[width=\linewidth]{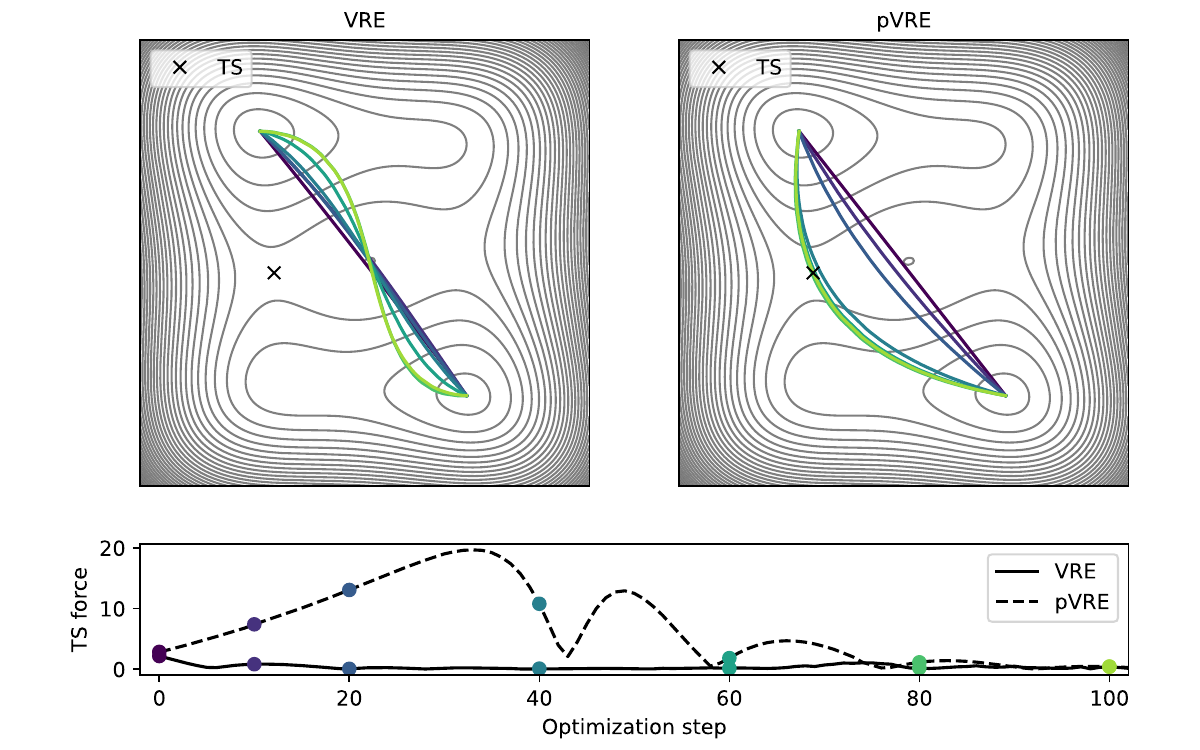}
    \caption{\textit{Comparison between the \vre{} and \pvre{} losses on the Wolfe potential.} Optimizing against \vre{} (top left) converges the path incorrectly to the gradient descent path from the maximum on the PES, while using \pvre{} (top right) converges the path to going through the TS. The bottom row shows the TS force during optimization. The colored lines and circles are corresponding measurements at the same time.}
    \label{fig:wolfe_vre_pvre}
\end{figure}

The \vre{} loss is only guaranteed to find the gradient descent path from any stationary point to which it converges. This is because the \vre{} path can converge onto a high-order saddle point or even a maximum, instead of the TS. This issue is more pronounced on a high-dimensional PES with many stationary points of various orders. To demonstrate this failure mode on a low-dimensional system, we show that the \vre{} path converges to the PES maximum of the Wolfe potential, in Fig.~\ref{fig:wolfe_vre_pvre}.

\subsection{The Projected Variational Reaction Energy (pVRE) Loss\label{app:pvre}}

The projected variational reaction energy (pVRE) loss is most affected by the highest energy point, thus targeting the TS, and better aligns with the assumption of TS theory.
The pVRE loss, reproduced here, is
\begin{equation}
    \pvremm = \int_{0}^{1} | \mathbf{F}(\mathbf{x}_\theta(t)) \cdot \mathbf{v}_\theta(t) | \, dt \tag{\ref{eq:pvre}} = \sum_a 2E_{a,TS} - E_{a,R} - E_{a,P} ,
\end{equation}
where $a$ sums over the maxima along the path, $E_R$ is the corresponding reactant energy, and $E_P$ is the corresponding product energy.
\pvre{} can be understood as the total variation of the energy profile, i.e., the summation of energy differences along \rxn{}.
Define the energy profile along the path as $E(\mathbf{x}_\theta(t))$.
Since
\begin{equation}
    \frac{d E(\mathbf{x}_\theta(t))}{dt}
    =
    \frac{d E(\mathbf{x}_\theta(t))}{d \mathbf{x}_\theta(t)} \cdot \frac{d \mathbf{x}_\theta(t)}{d t}
    =
    -\mathbf{F}(\mathbf{x}_\theta(t)) \cdot \mathbf{v}_\theta(t),
    \label{eq:energy_derivative}
\end{equation}
the pVRE loss can be written as the total variation of the energy profile
\begin{equation}
    \pvremm
    =
    \int_0^1
    \left|
    \frac{d E(\mathbf{x}_\theta(t))}{dt}
    \right|
    \, dt .
\end{equation}

A \pvre{} optimized path will prefer a single TS, generating a smooth path that avoids substructures and multi-step reactions that can be challenging for optimization methods to navigate.
We demonstrate this in the Muller-Brown potential (Fig.~\ref{fig:MB_vre_pvre}) and the Wolfe potential (Fig.~\ref{fig:wolfe_vre_pvre}), where \rxn{} targets the highest TS and avoids the others.
This behavior is valuable since the largest barrier (TS) corresponds to the reaction's rate-determining step, even for a multi-step reaction.
If one is interested in the elementary reactions, they can still be retrieved through IRCs from the TS \cite{Schlegel2011}.


This preference for transition-state regions can be seen directly from the loss gradient:
\begin{equation}
    \frac{\partial}{\partial \theta} \pvremm
    =
    \int_0^1
    \operatorname{sgn}\!\left(\frac{d E(\mathbf{x}_\theta(t))}{dt}\right)
    \frac{\partial}{\partial \theta}
    \left(
    \frac{d E(\mathbf{x}_\theta(t))}{dt}
    \right)
    \, dt .
\end{equation}
Given that the path endpoints are fixed,
\begin{equation}
    \frac{\partial E(\mathbf{x}_\theta(0))}{\partial \theta} = \frac{\partial E(\mathbf{x}_\theta(1))}{\partial \theta} = 0,
\end{equation}
and integrating by parts gives
\begin{equation}
    \frac{\partial}{\partial \theta} \pvremm
    =
    -
    \int_0^1
    \left[
    \frac{d}{dt}
    \operatorname{sgn}\!\left(\frac{d E(\mathbf{x}_\theta(t))}{dt}\right)\right]
    \frac{\partial E(\mathbf{x}_\theta(t))}{\partial \theta} 
    \, dt .
\end{equation}
The derivative of the sign function is a Dirac delta distribution,
\begin{equation}
    \frac{d}{dt}
    \operatorname{sgn}\!\left(\frac{d E(\mathbf{x}_\theta(t))}{dt}\right)
    =
    2
    \frac{d^2 E(\mathbf{x}_\theta(t))}{dt^2}
    \delta_{\mathrm{D}}
    \left(
    \frac{d E(\mathbf{x}_\theta(t))}{dt}
    \right),
\end{equation}
where $\delta_{\mathrm{D}}$ denotes the Dirac delta. Therefore, the gradient
is nonzero only where
\begin{equation}
    \frac{d E(\mathbf{x}_\theta(t))}{dt} = 0,
\end{equation}
i.e., at stationary points of the energy profile along \rxn{}. For
stationary points $t_a$, this can be written as
\begin{equation}
    \frac{\partial}{\partial \theta} \pvremm
    =
    -
    2
    \sum_a
    \operatorname{sgn}
    \left(
    \frac{d^2 E(\mathbf{x}_\theta(t_a))}{dt^2}
    \right)
    \frac{\partial E(\mathbf{x}_\theta(t_a)))}{\partial \theta}.
\end{equation}
If \rxn{} has only one energy barrier, the sum would only have one concave stationary point, which is the proposed TS, and can be simplified as
\begin{equation}
    \frac{\partial}{\partial \theta} \pvremm
    =
    2 \frac{\partial E(\mathbf{x}_\theta(t^*))}{\partial \theta},
\end{equation}
where $0<t^*<1$ is the proposed TS time along \rxn{}.
Using
\begin{equation}
    \frac{\partial E(\mathbf{x}_\theta(t))}{\partial \theta}
    =
    J_\theta(t)^\top
    \frac{\partial E(\mathbf{x}_\theta(t))}{\partial \mathbf{x}_\theta(t)}
    =
    -
    J_\theta(t)^\top
    \mathbf{F}(\mathbf{x}_\theta(t)),
\end{equation}
where $J_\theta(t)=\partial \mathbf{x}_\theta(t)/\partial\theta$ defined in \ref{eq:jacobian}, we obtain
\begin{equation}
    \frac{\partial}{\partial \theta} \pvremm
    =
    -2
    J_\theta(t^*)^\top
    \mathbf{F}(\mathbf{x}_\theta(t^*)).
    \label{eq:pvre_grad}
\end{equation}

Therefore, \pvre{} removes all maxima and minima along \rxn, and the TS at $t^*$ is the only contributing time point to the loss.
For interior maxima, the sign of Eq.~\ref{eq:pvre_grad} lowers the PES energy.
For interior minima, the sign of Eq.~\ref{eq:pvre_grad} raises the PES energy.
Therefore, \rxn{} is optimized to avoid any interior maxima or minima.
However, the TS cannot be removed, and thus it becomes the only contributing extremum to minimizing \pvre.
This provides a complementary view of why pVRE focuses optimization on transition-state structure: flat or monotonic portions of the path do not contribute to the first-order signal, whereas energy extrema contribute through delta-function weights.

This behavior is further intuited by the fact that \rxn{} accumulates no loss when traveling on an isoenergy line.
For \rxn{} to traverse an isoenergy line, its velocity must be perpendicular to the force, and by Eq.~\ref{eq:pvre} the integrand is 0.
The ability for \rxn{} to freely travel along isoenergy paths aligns with the statistical nature of chemical reactions, where structures of the same energy are equally likely to occur and thus \rxn{} should be able to freely travel between them.

\subsection{A Combined Variational Loss: Variational Error}
The work of Birkholz and Schlegel~\cite{Birkholz2015} defines a non-negative variational error loss to find the steepest descent reaction path.
The variational error
\begin{equation}
\varepsilon_\text{VRE} = \vremm - \pvremm
\label{eq:VRE_error}
\end{equation}
is the difference between the VRE and pVRE losses.
Minimizing this loss extremizes the dot product of \pvre: aligning the gradient of the PES with the velocity.
However, all gradient descent paths minimize $\varepsilon_\text{VRE}$; in fact, Eq.~\ref{eq:VRE_error} is independent of the stationary point.
That is, one can reach $\varepsilon_\text{VRE}=0$ by going over any feature where the PES gradient is 0: any PES peak, any saddle point, any PES minima.
Therefore, once \rxn{} corresponds to one of these extrema, there is likely no way to optimize \rxn{} towards the correct TS.
Instead, the path will be optimized to align \rxn{} with the PES gradient while staying at this extreme.

\subsection{Losses for Path Initialization: Geodesic Loss\label{app:geo_loss}}

Properly initializing the reaction path is a crucial step (Fig.~\ref{fig:workflow}b), and doing so incorrectly can place \rxn{} in a PES region where it will be optimized to the incorrect TS.
That is, the path initialization effectively confines the optimization to a PES region and must therefore be sensitive to the region with the correct TS.
This is necessary for multiple reasons: often accurate optimizations can be more expensive and cannot optimize over large PES regions; MLIPs 
can become undefined in some PES regions; and it is generally prohibitively challenging or expensive to optimize \rxn{} across many large PES structures to a new PES region on complex PES surfaces using especially ab initio methods.
Therefore, a quality path initializer must leverage chemical biases to select the PES region, balance PES evaluation cost, and often smooth the PES to allow easier optimization. 
Hence, one often uses a pairwise potential (Eq.~\ref{eq:pairpotential}) that is fast to evaluate, smooths the PES in a principled way, and still maintains chemical expectations.

Optimizing the initial path over the pairwise PES with \pvre{} and \vre{} suffers from catastrophic failures caused by approximations in the pairwise PES.
For \pvre{} this stems from the fact that on the pairwise PES the reactants and products are generally not at the minima, and there is often no TS in between.
Since Eq.~\ref{eq:pvre} has a monotonic integrand, the integral would be a constant $E(\mathbf{x}_\theta(1)) - E(\mathbf{x}_\theta(0))$ regardless of the path.
Thus \pvre{} is insensitive to the initial path in the common scenarios where reactant, product, and TS stationary points are missing and therefore cannot optimize the path.
In contrast, the \vre{} path is sensitive to the entire path, instead of a few special points like \pvre.
However, the total energy is not very indicative since the lengthening of one bond will be canceled by the shortening of another bond for a pairwise potential like Eq.~\ref{eq:pairpotential}.
This makes \vre{} insensitive to initial path changes.
Additionally, the energy can be reduced by compensating one bond motion with the opposite motion from another bond; this increases the possibility of visiting an unlikely structure where DFT and MLIPs may become undefined.

\begin{figure}[htb!]
    \centering
    \includegraphics[width=\linewidth]{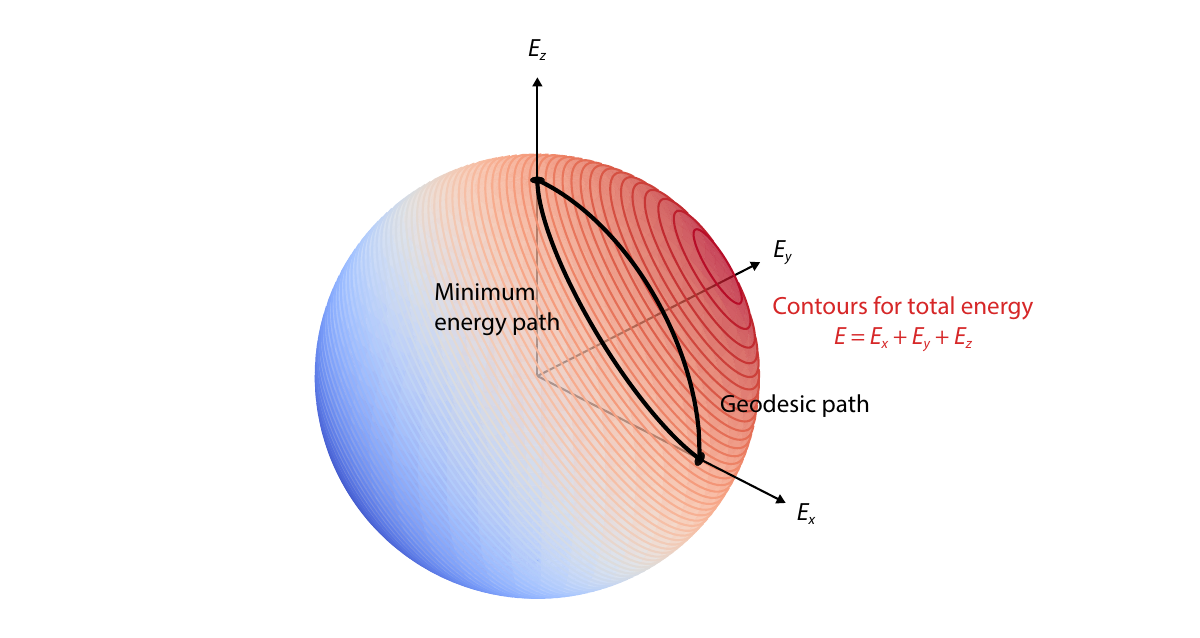}
    \caption{\textit{Comparison between the geodesic path and the minimum energy path.} Let a hypothetical PES has three energy contributions $E_x$, $E_y$, and $E_z$, and the feasible manifold is a sphere. The path with the smallest barrier, which is barrier-less in this case, lies on the plane $x+y+z=1$ and follows the energy contour. The geodesic, i.e., the path with the shortest arc, lies on the plane $y=0$.}
    \label{fig:geodesic}
\end{figure}

The geodesic loss (\geo) \cite{Zhu2019} can be intuited as minimizing the amount of displacement atoms must make while being rearranged from the reactant to the product structure.
This imposes a chemical intuition that the optimal reaction path can generally be optimized from the ``least-motion" or the spatially shortest path.
Instead of summing the pairwise energy contributions before evaluating the force or the loss integrand, \geo{} squares these individual pairwise contributions to minimize them separately
\begin{equation}
    \mathcal{L}_\text{geo} 
    = \int_{0}^{1} \sqrt{ \sum_{q \in \{(i,j)|i < j\}} \left| \mathbf{F}_q(\mathbf{x}_\theta(t)) \cdot \mathbf{v}_\theta(t) \right|^2 } \, dt
    = \int_{0}^{1} \left\| \mathbf{F}^\text{(pairs)}(\mathbf{x}_\theta(t)) \cdot \mathbf{v}_\theta(t) \right\| \, dt ,
    \tag{\ref{eq:geodesic}}
\end{equation}
Figure~\ref{fig:geodesic} demonstrates the difference between paths optimized by \geo{} and \pvre{} on a 3-dimensional example.
The path minimizing \geo{} does not correspond to the path crossing the minimal barrier minimizing \pvre{}.

To demonstrate the correspondence between \geo{} and \pvre, we derive \geo{} from \pvre.
To see this, consider a generalized pVRE loss as a sum of the pVRE losses from each pairwise energy term,
\begin{align}
    \mathcal{L}_\text{geo}^{(1)} 
    &= \sum_{q \in \{(i,j)|i < j\}} \mathcal{L}_{\text{pVRE},q} \\
    &= \sum_{q \in \{(i,j)|i < j\}} \int_{0}^{1} | \mathbf{F}_q(\mathbf{x}_\theta(t)) \cdot \mathbf{v}_\theta(t) | \, dt \\
    &= \int_{0}^{1} \sum_{q \in \{(i,j)|i < j\}} | \mathbf{F}_q(\mathbf{x}_\theta(t)) \cdot \mathbf{v}_\theta(t) | \, dt \\
    &= \int_{0}^{1} \left\| \mathbf{F}^\text{(pairs)}(\mathbf{x}_\theta(t)) \cdot \mathbf{v}_\theta(t) \right\|_1 \, dt,
    \label{eq:geodesic1}
\end{align}
where $\|\cdot\|_1$ is the 1-norm of a $(N(N-1)/2)$-dimensional vector, and $\mathbf{F}^\text{(pairs)} \in \mathbb{R}^{3N \times \frac{N(N-1)}{2}}$ is a matrix by stacking pairwise forces $\mathbf{F}_q \in \mathbb{R}^{3N}$, instead of summing them so that we can treat all pairs independently. Physically speaking, if one bond remains unchanged from the reactant to the product, this should be reflected in the learned path, and the contribution to the integral in Eq.~\ref{eq:geodesic1} should be zero.
This $\mathcal{L}_\text{geo}^{(1)}$, as a generalized \pvre{}, can still be independent of the path on the pairwise potential energy surface. To add a path dependence, we can replace the 1-norm with the 2-norm, i.e., the Euclidean norm, in a sense that one replaces the displacement with the distance or replaces the velocity with the speed, 
\begin{align}
    \mathcal{L}_\text{geo}^{(2)} 
    &= \int_{0}^{1} \sqrt{ \sum_{q \in \{(i,j)|i < j\}} \left| \mathbf{F}_q(\mathbf{x}_\theta(t)) \cdot \mathbf{v}_\theta(t) \right|^2 } \, dt \\
    &= \int_{0}^{1} \left\| \mathbf{F}^\text{(pairs)}(\mathbf{x}_\theta(t)) \cdot \mathbf{v}_\theta(t) \right\|_2 \, dt ,
    \label{eq:geodesic2}
\end{align}
which leads us to the geodesic loss \geo{} in Eq.~\ref{eq:geodesic}, to find the shortest path from the reactant to the product in this $N(N-1)/2$-dimensional space of energy terms. We note that Hait et al.~\cite{Hait2025} have applied the geodesic formalism to acquire the pVRE loss, without deeper physical explanation, but we find the opposite direction of understanding from \pvre{} to \geo{} is more physically intuitive.

\subsection{Transition State Losses}

\ppcrn{} is equipped with TS losses to speed up path converge \cite{Birkholz2015} and improve upon the TS structure.
Currently, these losses fall into two categories: the TS loss (\tsloss) and the TS region loss (\tsrloss)
\begin{align}
    \tslossmm &= \ell_\text{TS}(t_\text{TS}) \\
    \tsrlossmm &= \int_{t_i}^{t_f} \ell_\text{TS}(t) \, dt ,
\end{align}
where $\ell_\text{TS}$ is any user-defined function (including those used to optimize \rxn) applicable to the TS or TS region, $t_\text{TS}$ is the TS time, and $t_i < t_\text{TS} < t_f$ defines a region of the path around the TS.
More obvious choices for $\ell_\text{TS}$ that are supported by \ppcrn{} include the potential energy and the magnitude of its gradient at the TS.
The latter loss helps optimize the path to align with the TS saddle point.

The \tsrloss{} can be practically useful during path optimization by lowering the energy or the PES gradient of the highest region in the path.
Birkholz and Schlegel~\cite{Birkholz2015} showed that using a TS loss during optimization helped the path converge much faster.
The \tsrloss{} also helps define the TS when using \pvre.
Specifically, when applying \pvre{} optimization, \rxn{} is not penalized for traversing isoenergy geometries. Therefore, many structures near the TS can have the same or similar energies and potentially confuse a TS searching scheme.
For example, selecting the TS based on energy can be challenging, as the deviation from the maximum energy may be small in the TS region.
By minimizing the energy in the TS region, the TS will become more distinguishable.

In this work, we ran tests with both \tsloss{} and \tsrloss, but ultimately we did not use them.
While the TS losses speed up the reaction path convergence~\cite{Birkholz2015}, they lead to inferior paths due to faster convergence.
Many of these paths corresponded to multi-step reactions.
Given our goal is to find single-step reaction TS, the TS losses were not well suited for our specific task.
We do believe that the TS losses can be well-suited for other tasks, where one is interested in multi-step reactions or in higher TS structure precision.
We also note that the TS losses can be turned on at or near the end of the \ppcrn{} optimization (Supplementary Note~\ref{sec:curriculum}) to avoid path optimization effects while still improving the TS itself.


%% file: S03_convergence.tex
\label{app:convergence}

For each optimization stage, Popcornn monitors the path-integrated gradient
of the loss with respect to the path parameters (Eq.~\ref{eq:pvre_grad}).
A stage is declared converged when
\begin{equation}
    \left\|\frac{\partial}{\partial \theta} \pvremm\right\|_2 < \tau
\end{equation}
for $p$ consecutive optimization steps, where $\tau$ is the convergence
threshold and $p$ is the patience parameter. If this condition is not met,
the stage terminates after the prescribed maximum number of optimizer
iterations.

The threshold $\tau$ is chosen to correspond to a target residual force at
the predicted TS. The residual is anything that is nonzero because TS should have zero forces, while in practice a small nonzero residual is commonly allowed. Let
\begin{equation}
    F_2^{\mathrm{target}}
\end{equation}
denote the desired force magnitude of $\|\mathbf{F}(\mathbf{x}(t^\ast))\|_2$ at the TS candidate on the optimized path.
We wish to stop the optimization according to Eq.~\ref{eq:pvre_grad} at 
\begin{equation}
    \left\|\frac{\partial}{\partial \theta} \pvremm\right\|_2 
    = 
    2 \left\|
    J_\theta(t^*)^\top \mathbf{F}(\mathbf{x}_\theta(t^*))\right\|_2 
    < 
    \tau
\end{equation}
Using the singular-value bound given by
\begin{equation}
\begin{split}
    \|\mathbf{J}_\theta(t^\ast)^\top \mathbf{F}(\mathbf{x}_\theta(t^*))\|_2^2
    &=
    \mathbf{F}(\mathbf{x}_\theta(t^*))^\top \mathbf{J}_\theta(t^\ast) \mathbf{J}_\theta(t^\ast)^\top \mathbf{F}(\mathbf{x}_\theta(t^*)) \\
    &\geq 
    \sigma_{\min}(\mathbf{J}_\theta(t^\ast))^2 \|\mathbf{F}(\mathbf{x}_\theta(t^*))\|_2^2
\end{split}
\end{equation}
we have
\begin{equation}
    \|\mathbf{J}_\theta(t^\ast)^\top \mathbf{F}(\mathbf{x}_\theta(t^*))\|_2
    \ge
    \sigma_{\min}(\mathbf{J}_\theta(t^\ast)) \,
    \|\mathbf{F}(\mathbf{x}_\theta(t^*))\|_2 .
\end{equation}

For \pvre, this bound yields the estimate 
\begin{equation}
    \tau > \left\|\frac{\partial}{\partial \theta} \pvremm\right\|_2
    \ge
    2 
    \sigma_{\min}(\mathbf{J}_\theta(t^\ast)) \,
    \|\mathbf{F}(\mathbf{x}_\theta(t^*))\|_2
\end{equation}
Thus, we bound the loss gradient using the singular value lower bound and our force target $F_2^{\mathrm{target}}$
\begin{equation}
    \tau
    :=
    2 \,
    \sigma_{\min}(\mathbf{J}_\theta(t^\ast)) \,
    F_2^{\mathrm{target}} .
\end{equation}
With the common $F_2^{\mathrm{target}}$ of 0.05 eV/Ang and the singular value $\sigma_{\min}(\mathbf{J}_\theta(t^\ast)) \approx 1$ is estimated by \ref{eq:singular_value}, we set $\tau = 0.1$ throughout this work.



While \rxn{} converges, the integration tolerance must be at most the same magnitude as $\tau$.
At convergence, the optimization should reach a steady state at the optimal \rxn{} and the gradient of the loss with respect to the gradients is 0.
As previously mentioned, we bound this by $\tau$, and therefore at convergence
\begin{equation}
    \left\|\frac{\partial}{\partial \theta} \pvremm\right\|_2 \approx \tau
\end{equation}
Since the adaptive quadrature tolerance is given by 
\begin{equation}
    \max \left(a_\mathrm{tol}, r_\mathrm{tol}\,\left\|\frac{\partial}{\partial \theta}\pvremm\right\|_2\right),
\end{equation}
we set $a_\mathrm{tol}=0$ so the integration convergence is dependent on the loss objective and its gradient.
This determines $r_\text{tol}$ as 
\begin{equation}
    r_\mathrm{tol} \left\|\frac{\partial}{\partial \theta}\pvremm\right\|_2
    \approx
    \tau
    \quad 
    r_\mathrm{tol} = 1.
\end{equation}
The integration tolerance is therefore matched to the stopping threshold:
small enough not to mask convergence, but loose enough to avoid unnecessary
quadrature refinement during convergence.

%% file: S04_GSM.tex
\ppcrn{} can apply dynamic weighting to the path integral loss in a curriculum learning approach.
Curriculum learning is a machine learning technique that dynamically alters the loss to emphasize specific contributions the user wishes the optimization to focus on \cite{bengio2009curriculum, soviany2022curriculumlearningsurvey}.
Heuristically, this is similar to the growing string method (GSM), which optimizes end of strings that are continuously growing from the initial and final points.
\ppcrn{} is equipped with a growing-string-like curriculum that initially increases/decreases the path loss contributions near to and far from the endpoints.
The curriculum ends by evenly weighting all path points.

\ppcrn{} is also capable of increasing or decreasing specific loss contributions at different points of the optimization.
For example, the \tsloss{} can be absent in the first 3/4 of the path optimization (avoiding convergence to an inferior path) and then increased at the end to improve the TS.
Similarly, paths optimized with \pvre{} converge faster and find better TS than \vre, but can ignore PES structures by traversing isoenergy lines.
This can make a \pvre{} path difficult to interpret and will not resolve any local minima.
To leverage the superior \pvre{} performance and create a more interpretable path that resolves local minima (intermediate equilibrium structures), one can add \vre{} to the \pvre{} loss near the end of the optimization.

%% file: S05_parallel_processing.tex
Adaptive numerical integration is necessary to leverage the continuous path \rxn.
When \rxn{} is evaluated on a fixed grid of reaction times, the loss depends only on the structures at the grid points, which act as discrete images.
These structures are set by both the shape of the path and its velocity profile, i.e., how quickly the path moves through each region.
Changing the velocity profile, ``re-timing'' the path, moves the images along the path without changing its shape.
By changing the velocity profile, images ``fall'' toward regions of low loss during optimization.
For $\mathcal{L}_E$, the images fall down the PES toward low-energy structures such as \xr{} and \xp.
NEB prevents this by adding fictitious spring forces that keep the images evenly spaced~\cite{Jonsson1998, Henkelman2000}, effectively fixing the velocity profile~\cite{Ramakrishnan2025}.

For \vre{} and \pvre, re-timing alone can localize the images at stationary points of the PES, such as \xr, \xp, and the TS, where the force vanishes.
Although the exact \vre{} and \pvre{} integrals are independent of the velocity profile, their fixed-grid sums are not because re-timing changes which structures sit at the grid points.
The optimizer can therefore lower the fixed-grid loss without improving the path by placing the grid points at stationary points while the path crosses the high-loss regions between them.
In the limit, \rxn{} rests at \xr{} and \xp{} for nearly the entire reaction time and crosses between them within a single grid interval.
The images then coincide at the endpoints, and the fixed-grid loss vanishes regardless of the barrier the path crosses.

Instead of fixing the velocity profile through a fictitious spring force, the adaptive integrator adds evaluation points where the integration error is large and removes them where it is small, until the error falls below a set tolerance.
Where the path moves quickly, the integrand generally varies rapidly; the integrator adds points in these places, so re-timing no longer lowers the loss.
For \vre{} and \pvre, adaptive integration therefore removes the need for spring forces and keeps the loss within a set tolerance of the exact integral, so the gradient depends on the shape of the path rather than its velocity profile.

\subsection{\pdq}
Standard numerical integrators used for machine learning tasks are generally designed for ODEs~\cite{torchdiffeq} and require sequential evaluations in time, which would make \ppcrn{} computationally intractable.
Instead, we use \pdq~\cite{Hegazy2025}: an adaptive numerical quadrature method designed to parallelize path integration on GPUs with backpropagation compatibility.
\pdq, as an adaptive numerical quadrature method, divides the integral into steps that are evaluated and refined in parallel.
Steps are refined depending on their error ($\varepsilon_\text{step}$).
When given a new integral, \pdq{} will simultaneously evaluate integration steps based on a randomly sampled grid, shown as black points in Fig.~\ref{fig:model}b.
Regions with an error larger than the tolerance will be split in half, and new evaluation points will be made and evaluated in parallel (gray points in Fig.~\ref{fig:model}b).
This procedure continues until all integration steps satisfy the error criteria (white points in Fig.~\ref{fig:model}b) or have been split a maximum number of times (5 by default; Algorithm~\ref{alg:quadrature}).
The error criteria 
\begin{equation}
    \varepsilon_\text{step} < \mathrm{Max}(a_\text{tol}, r_\text{tol}\times\mathcal{L}')  ,
    \label{eq:pdq_error}
\end{equation}
where $a_\text{tol}$ and $r_\text{tol}$ are the absolute and relative tolerances, and $\mathcal{L}'$ is the latest path integral value.
The $a_\text{tol}$ and $r_\text{tol}$ values determine the integrators accuracy.
To evaluate an integration step's error Eq.~\ref{eq:pdq_error}, the integral ($\mathcal{L}'$) must be known.
\pdq{} defines $\mathcal{L}'$ as the integral with the current integration steps.
As the integration steps are refined, $\mathcal{L}'$ is updated with each parallel evaluation.
Thus, the integral value and the grid refinement converge simultaneously.

To further increase speed when integrating the same function, \pdq{} uses the previous evaluation points as an initial evaluation for the next optimization step.
\pdq{} avoids evaluating excess temporal points by pruning the evaluation time points at the end of each integration.
Note that \pdq{} uses all evaluated integration steps in the returned integral value; this pruned grid is only used to initialize the integration steps in the next optimization.
When initializing the integral with the previous integration times, \pdq{} generally requires 2 or 3 adaptive evaluations in the systems we tested.

\ppcrn{} is designed to handle systems that require large memory budgets, which may stem from expensive potentials or from a many-atom system.
To do this, \pdq{} calculates the maximum number of evaluations that can fit onto the GPU, and as a redundancy, reduces the evaluation batch size on out-of-memory errors.
\pdq{} can reduce the batch size to a single evaluation, thus handling any system where a single evaluation can fit within the GPU's memory.
The user may also set the evaluation batch size themselves to optimize \ppcrn{} on their own hardware.

%% file: S06_neb.tex
\label{app:neb}

Setting up the reference NEB and GSM methods is, in general, difficult and often case dependent.
We aim to make the comparison as fair as possible under a fixed set of hyperparameters. 
Below are some of the most important discretization and initialization hyperparameters we sampled.
All others are kept at default in ASE 3.28.0.

\begin{figure}[H]
    \centering
    \includegraphics[width=0.95\linewidth]{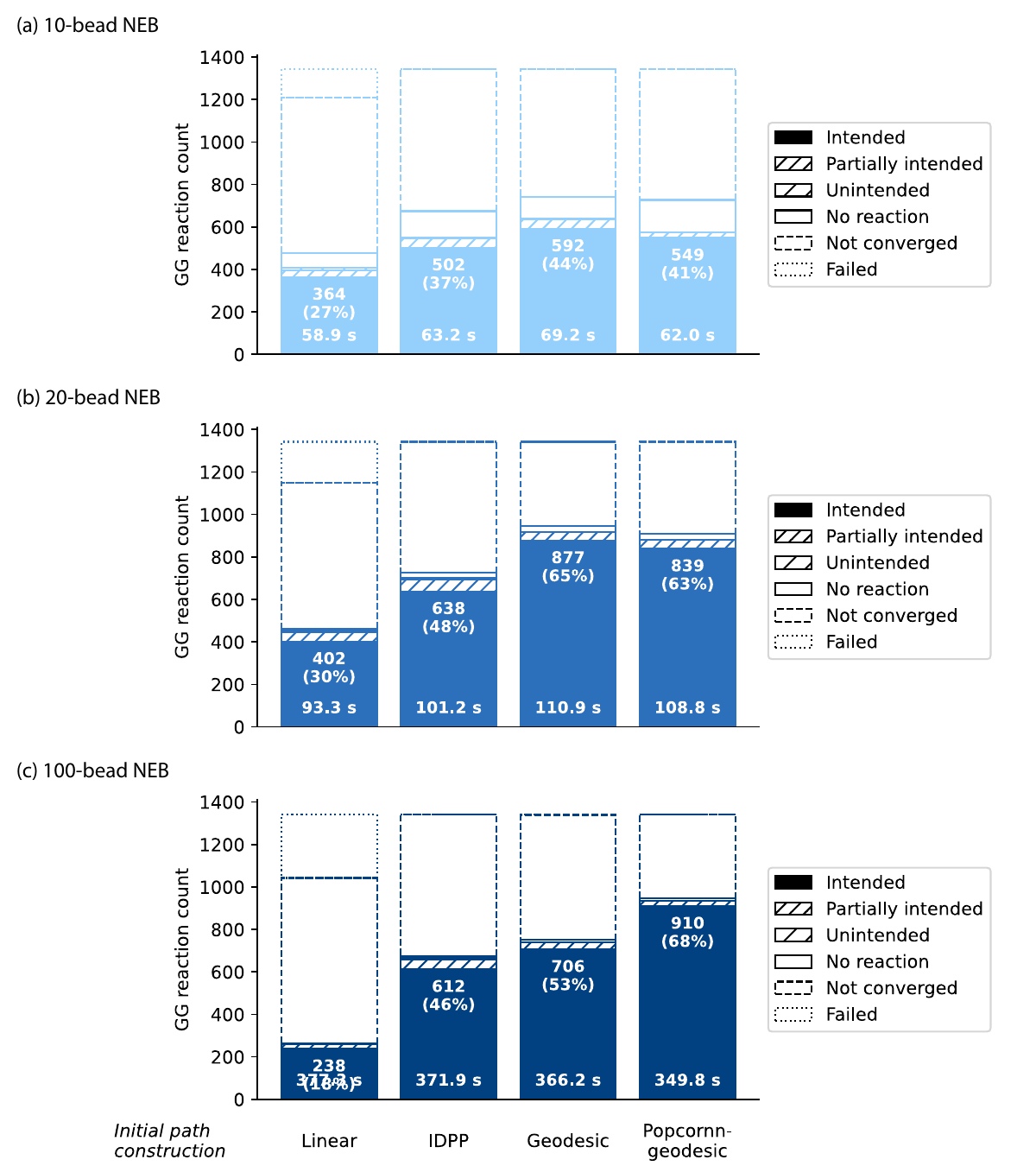}
    \caption{\textit{NEB setting screening on Grambow-Green out-of-distribution reactions.} We evaluate NEB's performance with different hyperparameters. We explore the number of beads: (a) 10 beads, (b) 20 beads, and (c) 100 beads. For each number of beads, we vary the initial path construction method (linear interpolation, IDPP interpolation, discretized geodesic interpolation, and \ppcrn-geodesic). The number at the bottom of each bar is the average time, in seconds, for the TS optimization to converge.}
    \label{fig:nebparam_gg}
\end{figure}



%% file: S09_pseudocode.tex
\label{app:algorithms}

This Supplementary Note gives pseudocode for \ppcrn{} as implemented in the code repository~\cite{hegazy2026popcornn}; we checked it against release 0.2.1 of the code (commit \texttt{15dd905}, 15 November 2026) and release 1.4.1 of the \pdq{} implementation \texttt{padaquad}.
Algorithm~\ref{alg:popcornn} is a complete \ppcrn{} run, and Algorithms~\ref{alg:path_eval}--\ref{alg:ts_search} expand its steps.
Settings stated in Methods appear here as inputs.
Settings not stated in Methods take the code defaults: the patience $p = 1$, the number of output structures $n_\text{rec} = 101$, the random seed $0$, single-precision (32-bit) arithmetic, the maximum number of times an integration step may be split $M_\text{split} = 5$, and the error norm $\|\cdot\|_\nu$ with $\nu = 2$ (the Euclidean norm of a vector).
Throughout, $N$ is the number of atoms, $\theta$ denotes the parameters of the path network \fnn{} with $D$ entries in total, and $t \in [0, 1]$ is the reaction time.
Derivatives are computed by automatic differentiation, which obtains exact derivatives of a computer program by applying the chain rule to each of its operations; reverse-mode automatic differentiation (backpropagation) gives the derivative of a scalar with respect to all of $\theta$ in one pass.

\subsection{A Complete \ppcrn{} Run\label{app:alg_overview}}

A \ppcrn{} run is a sequence of optimization stages that share the path parameters $\theta$.
For the molecular benchmarks in this work there are two stages, initial path construction and path optimization on the MLIP (Methods), each with its own potential, integrand, and Adam optimizer.
Every optimizer iteration integrates the gradient of the path loss along the whole path (Algorithm~\ref{alg:path_gradient}) and updates all of $\theta$ at once.
After the last iteration of a stage, one more integration records $dE/dt$ along the path, from which the TS is located (Algorithm~\ref{alg:ts_search}); the TS that \ppcrn{} reports is the one found after the final stage.
Optionally, the loss integral and the TS search are also computed at every iteration for monitoring; they do not change $\theta$.

\begin{algorithm}[htbp]
\caption{A complete \ppcrn{} run.}\label{alg:popcornn}
\begin{algorithmic}[1]
\Require reactant and product structures $\xrmm, \xpmm \in \mathbb{R}^{3N}$, relaxed and aligned (Methods); optional intermediate structures $\mathbf{x}_1, \dots, \mathbf{x}_K$; stages $s = 1, \dots, S$, each with a potential $U_s$, an integrand $\ell_s$, absolute and relative quadrature tolerances $a_{\text{tol},s}$ and $r_{\text{tol},s}$, a learning rate $\eta_s$, an optional convergence threshold $\tau_s$, a patience $p_s$, and a maximum number of iterations $M_s$
\Ensure the path at $n_\text{rec}$ evenly spaced times, the TS estimate $\mathbf{x}_\theta(t^*)$, and for each stage whether its convergence test was met
\State Set the random seed; initialize $\theta$ with PyTorch's default initialization for linear layers
\If{intermediate structures are given} \Comment{not used in this work}
    \State $t_k \gets k/(K+1)$ for $k = 0, \dots, K+1$, with $\mathbf{x}_0 = \xrmm$ and $\mathbf{x}_{K+1} = \xpmm$
    \State Fit $\theta$ with Adam (learning rate $10^{-3}$) to minimize $\frac{1}{3N(K+2)}\sum_{k=0}^{K+1} \|\mathbf{x}_\theta(t_k) - \mathbf{x}_k\|^2$, stopping when it falls below $10^{-8}$ or after $10^{4}$ iterations
\EndIf
\For{$s = 1, \dots, S$}
    \State Attach $U_s$ to the path; create a new Adam optimizer with learning rate $\eta_s$; $c \gets 0$ \Comment{$c$: consecutive iterations below $\tau_s$}
    \For{$m = 1, \dots, M_s$}
        \State $\mathbf{G} \gets \Call{PathGradient}{\theta, U_s, \ell_s, a_{\text{tol},s}, r_{\text{tol},s}}$ \Comment{Algorithm~\ref{alg:path_gradient}; $\mathbf{G} \approx \partial \mathcal{L}_s / \partial \theta$}
        \If{$\tau_s$ is set \textbf{and} $\|\mathbf{G}\|_\nu < \tau_s$}
            \State $c \gets c + 1$
        \Else
            \State $c \gets 0$
        \EndIf
        \If{$c \geq p_s$}
            \State \textbf{break} \Comment{converged (Supplementary Note~\ref{app:convergence})}
        \EndIf
        \State $\theta \gets \Call{AdamStep}{\theta, \mathbf{G}, \eta_s}$ \Comment{an optional learning-rate schedule also steps here}
    \EndFor
    \State $(\mathbf{G}, \mathcal{Q}) \gets \Call{PathGradient}{\theta, U_s, \ell_s, a_{\text{tol},s}, r_{\text{tol},s}}$, recording the samples $\mathcal{Q}$ of $dE/dt$ (Algorithm~\ref{alg:path_gradient}); also integrate the loss for reporting, with its own tolerances ($a_\text{tol} = 10^{-4}$, $r_\text{tol} = 0$)
    \State $t^* \gets \Call{TSSearch}{\mathcal{Q}}$ \Comment{Algorithm~\ref{alg:ts_search}}
\EndFor
\State \Return $\mathbf{x}_\theta(t_i)$ for $t_i = (i-1)/(n_\text{rec}-1)$, $i = 1, \dots, n_\text{rec}$, and $\mathbf{x}_\theta(t^*)$
\end{algorithmic}
\end{algorithm}

\subsection{Evaluating the Path\label{app:alg_path}}

Algorithm~\ref{alg:path_eval} evaluates the path model of Eq.~\ref{eq:path} at a batch of reaction times, with the network input rescaled to $\tilde t = 2t - 1$ (Supplementary Note~\ref{app:path_architecture}).
All quantities are computed for the whole batch at once and stay differentiable with respect to $\theta$.
The velocity \vel{} is the derivative of the path with respect to $t$, obtained by automatic differentiation rather than by finite differences.
The potential is called once per batch; the repository provides one wrapper per supported MLIP that returns energies and atomic forces for a batch of structures.
For the force-based losses, the gradient with respect to $\theta$ passes through the forces, so it needs the derivative of the forces with respect to the atomic positions.
For models whose forces are the negative energy gradient, this derivative is the negative of the PES Hessian (the matrix of second derivatives of the energy), and automatic differentiation applies it to vectors without forming the full matrix.
For models that predict forces directly, or for the analytic forces of the pair potential, automatic differentiation differentiates that force expression instead.

\begin{algorithm}[htbp]
\caption{Evaluating the path at a batch of reaction times.}\label{alg:path_eval}
\begin{algorithmic}[1]
\Function{EvaluatePath}{$t_1, \dots, t_n$; $\theta$, $U$}
    \For{all $j = 1, \dots, n$ in parallel}
        \State $\mathbf{x}_j \gets (1 - t_j)\, \xrmm + t_j\, \xpmm + t_j (1 - t_j)\, \mathbf{f}_\theta(2t_j - 1)$ \Comment{Eq.~\ref{eq:path}}
        \State Set the network perturbation of atoms fixed by constraints to zero
        \State $\mathbf{v}_j \gets d\mathbf{x}_j / dt$, by automatic differentiation
        \State For periodic cells, wrap $\mathbf{x}_j$ into the cell before calling the potential \Comment{not used in this work}
        \State $(E_j, \mathbf{F}_j) \gets U(\mathbf{x}_j)$ \Comment{$\mathbf{F}_j = -\nabla E_j$ for conservative models}
        \State For the pair potential (Eq.~\ref{eq:pairpotential}), also return the pairwise forces $\mathbf{F}^\text{(pairs)}_j \in \mathbb{R}^{3N \times N(N-1)/2}$
    \EndFor
    \State \Return those of $\{\mathbf{x}_j, \mathbf{v}_j, E_j, \mathbf{F}_j, \mathbf{F}^\text{(pairs)}_j\}_{j=1}^{n}$ that the caller asks for
\EndFunction
\end{algorithmic}
\end{algorithm}

\subsection{Gradient of the Path Loss\label{app:alg_gradient}}

Algorithm~\ref{alg:path_gradient} computes the parameter gradient by integrating the gradient of the integrand, $\int_0^1 \partial \ell / \partial \theta \, dt$ in Eq.~\ref{eq:grad_path_integral}, as described in Results; this is the code default, and the other option, differentiating the completed integral of $\ell$, is not used in this work.
The integrand passed to the adaptive quadrature is therefore the vector $\partial \ell(t) / \partial \theta \in \mathbb{R}^{D}$.
For a batch of $n$ times, the $n$ gradient vectors come from one batched reverse-mode pass.
When requested, the same evaluations also record $dE/dt = -\mathbf{F} \cdot \mathbf{v}$ (Eq.~\ref{eq:energy_derivative}) at every evaluated time.
Only the samples in integration steps that the quadrature finally accepts are kept; samples in steps that it splits are discarded.
Algorithm~\ref{alg:ts_search} uses the kept samples.

\begin{algorithm}[htbp]
\caption{Gradient of the path loss with respect to the path parameters.}\label{alg:path_gradient}
\begin{algorithmic}[1]
\Function{PathGradient}{$\theta$, $U$, $\ell$, $a_\text{tol}$, $r_\text{tol}$}
    \Function{$g$}{$t_1, \dots, t_n$} \Comment{integrand: one gradient vector per time}
        \State Evaluate the path at $t_1, \dots, t_n$ \Comment{Algorithm~\ref{alg:path_eval}}
        \State $\ell_j \gets \sum_m w_m\, \ell_{m,j}$ over the configured terms; this work uses one term with $w = 1$: $|\mathbf{F}_j \cdot \mathbf{v}_j|$ (pVRE, Eq.~\ref{eq:pvre}), $\|\mathbf{F}_j\|\,\|\mathbf{v}_j\|$ (VRE, Eq.~\ref{eq:vre}), or $\|(\mathbf{F}^{\text{(pairs)}}_j)^{\top} \mathbf{v}_j\|$ (geodesic, Eq.~\ref{eq:geodesic})
        \State Record $\big(t_j,\, -\mathbf{F}_j \cdot \mathbf{v}_j\big)$ in $\mathcal{Q}$ \Comment{only when samples are requested}
        \State $\mathbf{g}_j \gets \partial \ell_j / \partial \theta$ for all $j$, by batched reverse-mode automatic differentiation
        \State \Return $[\mathbf{g}_1, \dots, \mathbf{g}_n] \in \mathbb{R}^{n \times D}$
    \EndFunction
    \State $\mathbf{G} \gets \Call{AdaptiveQuadrature}{g, a_\text{tol}, r_\text{tol}}$ \Comment{Algorithm~\ref{alg:quadrature}}
    \State Keep in $\mathcal{Q}$ only the samples in accepted integration steps, and sort them by time
    \State \Return $\mathbf{G}$ (and $\mathcal{Q}$)
\EndFunction
\end{algorithmic}
\end{algorithm}

\subsection{Adaptive Quadrature\label{app:alg_quadrature}}

Algorithm~\ref{alg:quadrature} gives the parallel adaptive quadrature of \pdq{} as described in Supplementary Note~\ref{app:pdq_parallelism} and Ref.~\cite{Hegazy2025}.
\ppcrn{} calls it through the \texttt{padaquad} package, an implementation of \pdq{} that is not part of the \ppcrn{} repository; this algorithm follows that source.
The integration interval is divided into integration steps.
Each step carries an integral estimate and an error estimate from an embedded rule pair: two quadrature rules of different order that share evaluation points, whose difference estimates the error.
\ppcrn{}'s default, used in this work, is the 7-point Gauss--Kronrod rule with its embedded 3-point Gauss rule.
The integrand is evaluated at 9 points per step: the 7 Gauss--Kronrod points and the step's two ends, which have zero weight in both rules.
Because the integrand is the gradient vector, the integral estimate $\mathbf{G}'$ (the $\mathcal{L}'$ of Eq.~\ref{eq:pdq_error}) and the step errors are vectors in $\mathbb{R}^{D}$, and the error criterion compares their $\nu$-norms; the same norm is used for the convergence test in Algorithm~\ref{alg:popcornn}.
With $a_\text{tol} = 0$, as in this work, the criterion depends only on $r_\text{tol}$, except that a step is never required to be more accurate than floating-point rounding allows.
Steps accepted earlier are tested again against each new $\mathbf{G}'$, because $\mathbf{G}'$ changes as refinement proceeds.
A step that has already been split $M_\text{split}$ times is accepted even if it fails the criterion.
After the loop, every accepted step is tested once more against the final $\mathbf{G}'$; if any step fails, \texttt{padaquad} logs a warning and marks the integral as not converged, but \ppcrn{} uses the integral unchanged.
An integrand value that is not finite stops the run with an error.
Each \ppcrn{} stage creates a new integrator, so warm starting (Supplementary Note~\ref{app:pdq_parallelism}) and a reduced batch size carry over between the calls of a stage but not between stages.

\begin{algorithm}[htbp]
\caption{Parallel adaptive quadrature (\pdq).}\label{alg:quadrature}
\begin{algorithmic}[1]
\Function{AdaptiveQuadrature}{$g$, $a_\text{tol}$, $r_\text{tol}$}
    \If{this is the first call of the stage}
        \State New $\gets$ 11 integration steps: $[0, 1]$ is cut into 3 equal segments, and each segment is cut at 3 uniformly random times
        \State $B \gets$ the user's batch-size cap if set, else, before each round of evaluations, the largest number of evaluations that fits in the free GPU memory, predicted by a memory model fitted at the start of the call
    \Else \Comment{warm start}
        \State New $\gets$ the steps stored by the previous call
        \State $B \gets$ the batch size stored by the previous call
    \EndIf
    \State Accepted $\gets \emptyset$
    \Repeat
        \State Evaluate $g$ at the 9 points of every step in New, in batches of at most $B$ points
        \Statex \hspace{\algorithmicindent}\Comment{if GPU memory runs out, retry with $B \gets \max(1, \min(\mathrm{round}(0.75B),\, B - 1))$}
        \State For each step $k$ in New, compute its integral $\mathbf{I}_k$ and error $\boldsymbol{\varepsilon}_k$ from the embedded rule pair
        \State $\mathbf{G}' \gets \sum_{k \in \text{Accepted} \cup \text{New}} \mathbf{I}_k$ \Comment{current estimate of the integral}
        \State Failed $\gets \{k \in \text{Accepted} \cup \text{New} : \|\boldsymbol{\varepsilon}_k\|_\nu \geq \max(a_\text{tol},\, r_\text{tol} \|\mathbf{G}'\|_\nu,\, 50\,\epsilon\,\|\mathbf{I}_k\|_\nu)$ and step $k$ has been split fewer than $M_\text{split}$ times$\}$
        \Statex \hspace{\algorithmicindent}\Comment{Eq.~\ref{eq:pdq_error}; $\epsilon$: machine epsilon of the working precision}
        \State Accepted $\gets (\text{Accepted} \cup \text{New}) \setminus \text{Failed}$
        \State New $\gets$ the two halves of every failed step
    \Until{New is empty}
    \State Store $B$ and the accepted steps for the next call, after merging neighbouring pairs whose merged error, predicted from the integrand values already computed, is below $10\%$ of the tolerance and halving the steps that still fail because they reached $M_\text{split}$
    \State \Return $\mathbf{G}' = \sum_{k \in \text{Accepted}} \mathbf{I}_k$
\EndFunction
\end{algorithmic}
\end{algorithm}

\subsection{Locating the Transition State\label{app:alg_ts}}

Algorithm~\ref{alg:ts_search} is the full TS search summarized in Methods.
It uses the samples recorded during the final integration of a stage and brackets each zero of $dE/dt$ between neighbouring samples.
Because each integration step is evaluated at its two ends, the first and last samples lie exactly at the reactant and product ($t = 0$ and $t = 1$).
These two samples are excluded from the brackets because, at and close to the reactant and product minima, $dE/dt$ is near zero and its sign is unreliable.
The reactant and product times are always candidates, so that a path without an interior maximum returns the higher-energy endpoint.
Taking the highest-energy candidate also discards zeros of $dE/dt$ that are energy minima.

\begin{algorithm}[htbp]
\caption{Locating the TS along the optimized path.}\label{alg:ts_search}
\begin{algorithmic}[1]
\Require samples $\mathcal{Q} = \{(t_i, d_i)\}_{i=1}^{n}$ sorted by $t_i$, with $d_i = dE/dt$ at $t_i$
\Function{TSSearch}{$\mathcal{Q}$}
    \State $\mathcal{C} \gets \{0, 1\}$ \Comment{reactant and product times}
    \For{$i = 2, \dots, n-2$}
        \If{$d_i \, d_{i+1} \leq 0$ \textbf{and not} $d_i = d_{i+1} = 0$} \Comment{sign change of $dE/dt$}
            \State $\mathcal{C} \gets \mathcal{C} \cup \{\, t_i - d_i (t_{i+1} - t_i) / (d_{i+1} - d_i) \,\}$ \Comment{zero of $dE/dt$}
        \EndIf
    \EndFor
    \State Evaluate $E$ and $\mathbf{F}$ at every $t \in \mathcal{C}$ in one batched call \Comment{Algorithm~\ref{alg:path_eval}}
    \State $t^* \gets \arg\max_{t \in \mathcal{C}} E(\mathbf{x}_\theta(t))$
    \State \Return $t^*$, $E(\mathbf{x}_\theta(t^*))$, $\mathbf{F}(\mathbf{x}_\theta(t^*))$, and the barrier $E(\mathbf{x}_\theta(t^*)) - E(\xrmm)$
\EndFunction
\end{algorithmic}
\end{algorithm}

\subsection{Where Each Step Is Implemented\label{app:alg_code}}

Table~\ref{tab:alg_code} maps each algorithm to the module and function of the \ppcrn{} repository~\cite{hegazy2026popcornn} that implement it.
A new MLIP is added by writing one wrapper in \texttt{popcornn/potentials/} that returns energies and forces for a batch of structures, and adding its name to \texttt{get\_potential} in \texttt{popcornn/potentials/\_\_init\_\_.py}.

\begin{table}[htb]
\caption{Location of each algorithm in the \ppcrn{} repository (release 0.2.1) and in \texttt{padaquad} (release 1.4.1).}\label{tab:alg_code}
\begin{tabular}{@{}ll@{}}
\toprule
Algorithm & Module (function) \\
\midrule
\ref{alg:popcornn}, input processing & \texttt{popcornn/tools/images.py} (\texttt{process\_images}) \\
\ref{alg:popcornn}, seed and initial fit & \texttt{popcornn/popcornn.py} (\texttt{Popcornn.\_\_init\_\_}), \\
 & \texttt{popcornn/optimization/initialize\_path.py} (\texttt{initialize\_path}) \\
\ref{alg:popcornn}, stages and iteration loop & \texttt{popcornn/popcornn.py} (\texttt{Popcornn.optimize\_path}, \texttt{\_optimize}) \\
\ref{alg:popcornn}, Adam step and convergence & \texttt{popcornn/optimization/optimizer.py} (\texttt{PathOptimizer}) \\
\ref{alg:popcornn}, outputs & \texttt{popcornn/popcornn.py} (\texttt{get\_discrete\_path}, \texttt{get\_ts}) \\
\ref{alg:path_eval}, path model & \texttt{popcornn/paths/} (\texttt{mlp.py}, \texttt{linear.py}, \texttt{base\_path.py}) \\
\ref{alg:path_eval}, potentials & \texttt{popcornn/potentials/} (pair potential: \texttt{repel.py}) \\
\ref{alg:path_gradient}, integrands & \texttt{popcornn/tools/integrand.py} \\
\ref{alg:path_gradient}, gradient integration & \texttt{popcornn/tools/integrator.py} (\texttt{PathIntegrator}) \\
\ref{alg:quadrature}, adaptive quadrature & \texttt{padaquad.adaptive\_quadrature}, called from \texttt{integrator.py} \\
\ref{alg:ts_search}, TS search & \texttt{popcornn/paths/base\_path.py} (\texttt{BasePath.ts\_search}) \\
\botrule
\end{tabular}
\end{table}
\FloatBarrier